\documentclass[aip,jcp,preprint]{revtex4-1}

\usepackage[version=4]{mhchem}
\usepackage{graphicx}

\begin{document}

\title{Adsorbate phase transitions on nanoclusters from nested sampling}

\author{Thanawitch Chatbipho}
\affiliation{Department of Chemistry, University of Warwick, Coventry, CV4 7AL, UK}

\author{Ray Yang}
\affiliation{Department of Chemistry and Institute of Materials Science and Engineering, Washington University in St.\@~Louis, St.\@~Louis, MO, USA}

\author{Robert B.\@~Wexler}
\email[]{wexler@wustl.edu}
\affiliation{Department of Chemistry and Institute of Materials Science and Engineering, Washington University in St.\@~Louis, St.\@~Louis, MO, USA}

\author{Livia B.\@~P\'{a}rtay}
\email[]{Livia.Bartok-Partay@warwick.ac.uk}
\affiliation{Department of Chemistry, University of Warwick, Coventry, CV4 7AL, UK}

\date{\today}

\begin{abstract}

Nested sampling was employed to investigate adsorption equilibria on the truncated-octahedral Lennard--Jones nanocluster LJ$_{38}$ while systematically varying adsorbate--surface well depth and Lennard--Jones size parameters.
Evaluation of the canonical partition function over a wide temperature range identifies two successive phase transitions: (i) condensation of the gas phase onto the cluster surface at higher temperatures, and (ii) lateral rearrangement of the adsorbed layer at lower temperatures.
For identical interactions, the condensate first populates both three- and four-fold hollow sites; when adsorbate--adsorbate interactions are weakened, preference shifts to the four-coordinated (100) sites.
Size mismatch governs low-temperature behavior: smaller adsorbates aggregate to increase mutual contacts, whereas larger ones distribute more evenly to maximize coordination with the cluster.
These findings highlight key trends in facet competition and lattice mismatch, and showcase nested sampling as an automated, unbiased tool for exploring surface configurational space and guiding investigations of more complex, realistic interfaces.

\end{abstract}


\maketitle %\maketitle must follow title, authors, abstract and \pacs

\section{Introduction}

The structure and behavior of solid-fluid interfaces have been extensively studied, as nanoscale properties often differ significantly from those of bulk materials due to pronounced quantum and geometric size effects.
These surface-specific phenomena (such as adsorption, phase transitions in adsorbed layers, and interfacial reconstructions) underlie numerous processes, ranging from energy storage \cite{pomerantseva2019energy} and atmospheric chemistry \cite{buseck2008nanoparticles, rabajczyk2020metal} to heterogeneous catalysis \cite{liu2018metal} and biotechnology. \cite{bio-nano, bio-nano2, mitchell2021engineering}
Nanoparticles are of particular interest because their high surface-to-volume ratios and abundance of low-coordinated active sites not only enhance reactivity but can also afford practical advantages---for instance, ultra-fine Au/Fe-, Au/Co-, and Au/Ni-oxide composites fully oxidize CO at $-70 \ ^{\circ}\mathrm{C}$ while remaining stable in humid gas streams, a performance unattainable with bulk catalysts. \cite{haruta1987novel}
Moreover, characteristics like particle size, morphology, and composition strongly influence nanoparticle properties, enabling researchers to tune catalytic activity, thermal stability, optical response, and magnetic behavior for diverse applications. \cite{NP_medicine, NP_magnetic, NP_industrial, liu2018metal, NP_review, bio-nano2, baig2021nanomaterials}
Surface adsorption and reorganization are often critical in such applications.
For example, palladium catalysts exhibit facet-dependent activity in hydrodechlorination reactions. \cite{jiang2020mechanistic, facet_specific_adsorption}
Similarly, small gas molecules (\ce{H2}, \ce{O2}, \ce{CO}) bind preferentially to specific sites on Au-Rh bimetallic clusters \cite{AuRh_clusters} and on copper nanoparticle surfaces. \cite{cao2002static, Cu_cluster_sites}
Thus, to design nanoparticles with targeted functions, it is essential to understand both the thermodynamics and kinetics of their surface processes.

Computational modeling has been an indispensable tool for obtaining an atomic-scale understanding of surface composition and structure, as well as the interaction of surfaces with adsorbates. \cite{ferrando2008nanoalloys, bio-nanomaterials, calvo2015thermodynamics, pidko2017toward, douglas2021plasmonic}
These simulations support the interpretation of spectroscopic data, elucidate complex reaction pathways, and enable useful predictions for large-scale screening. \cite{liu2018metal, catlow2020computational}
Nanoparticle surfaces, however, pose a particular challenge for modeling.
Owing to the myriad possible surface terminations, reconstructions, and low-symmetry features (such as edges and defects), modeling efforts often focus on only a few high-symmetry terminations and ground-state structures, neglecting the full diversity of surface configurations and their stability at finite temperatures (including anharmonic effects) under conditions of practical relevance. \cite{grajciar2018towards}
Random structure search and global optimization techniques have been crucial in predicting stable nanocluster configurations \cite{wales1997global, lazauskas2017efficient, woodley2020structure, nabi2022ab} and comparing competing morphologies. \cite{baletto2002crossover, CuPt_ns, casey2021theoretical}
However, because these approaches typically consider only the 0~K potential energy surface (neglecting entropic effects), they are seldom applied to study adsorption phenomena under realistic conditions. \cite{borg2005density}

Low-temperature surface energies can be evaluated through the \textit{ab initio} atomistic thermodynamics approach pioneered by Scheffler and coworkers. \cite{reuter2001composition, stampfl2002catalysis, lee2024rise}
Moreover, Wexler, Qiu, and Rappe combined density functional theory with grand-canonical Monte Carlo (GCMC) simulations to accurately and automatically sample the phase space of oxide overlayers on Ag(111), reproducing important aspects of its surface phase diagram without pre-selecting candidate structures. \cite{wexler2019automatic}
This method provides valuable insights, but at a significant computational cost.
To improve efficiency, machine-learning strategies have likewise been employed to develop generalized adsorption models for nanocluster surfaces. \cite{MLadsorption_model}
Zhou, Scheffler, and Ghiringhelli introduced a replica-exchange grand-canonical sampling technique to estimate differences in free energies for the prediction of surface phase diagrams. \cite{zhou2019determining}
Their reliance on discrete temperature and chemical-potential grids can also lead to coarse sampling of phase space around surface phase transitions.
Here, we adopt an alternative nested sampling (NS) approach that avoids such coarse discretization by constructing a set of slabs equidistant in the natural logarithm of the surface configuration-space volume.

Recently, we demonstrated that NS can be employed to efficiently sample the configuration space of surfaces, yielding thermodynamically relevant adsorbate structures without requiring any prior knowledge of the stable phases. \cite{1st_NS_paper, yang2024surface} 
In that proof-of-concept study, we used NS to calculate coverage-temperature phase diagrams for adsorbates on four facets of a Lennard--Jones (LJ) face-centered cubic (fcc) solid.
From these NS simulations, we constructed the canonical partition function and computed ensemble-averaged thermodynamic properties, such as the constant-volume heat capacity and surface order parameters.
Notably, this approach revealed phase transitions on both flat and stepped facets, including an enthalpy-driven condensation at higher temperatures and an entropy-driven reordering at lower temperatures.
The surface NS method is general and can be applied not only to flat and stepped surfaces but also to arbitrary ``host'' structures for adsorption, including nanoparticles, clusters, channels, or porous materials.
To demonstrate this versatility, the present work applies surface NS to a model cluster with multiple facets: the fcc-truncated-octahedral global-minimum-energy structure of 38 LJ particles, LJ$_{38}$. \cite{bib:wales_basin_LJ}
We use this system to showcase the method's ability to capture the competition between adsorption sites of different coordination, as well as to examine the effects of lattice mismatch and varying adsorbate--surface interaction strengths, by changing the LJ length and energy parameters of the free particles adsorbing onto the surface.
We also compare the computational cost of our method to that of the widely used parallel tempering technique, which---provided that the temperature ladder contains sufficiently overlapping replicas---can facilitate crossing of energy barriers and thereby enhance exploration of configuration space.

\section{Computational Methods} \label{computational-methods}

\subsection{Surface Nested Sampling}

NS is an iterative Monte Carlo algorithm \cite{bib:skilling, bib:skilling2, NS_all_review} that contracts the accessible phase-space volume in equal logarithmic steps, thereby transforming the high-dimensional integral for the canonical partition function into a tractable one-dimensional sum. \cite{NS_mat_review}
Unlike importance-sampling methods that require \textit{a priori} knowledge of relevant phases, NS explores configuration space ``top-down'' and produces the partition function, and hence all equilibrium observables, directly.
Recent work provides derivations for bulk systems in the canonical \cite{1st_NS_paper, Frenkel_NS} and isobaric ensembles, \cite{pt_phase_dias_ns, ConPresNS} as well as the surface-specific modifications we introduced; \cite{yang2024surface} interested readers are referred there for details.

\subsubsection{Workflow Overview}

The following overview summarizes how we adapt the standard materials application of nested sampling to a fixed nanocluster substrate with freely mobile adsorbates.

\begin{enumerate}

\item \textbf{Initial live set:} A fixed $K$ number of ``live'' configurations (walkers) is drawn uniformly from the prior volume.

\textit{For cluster adsorption}, we treat the nanocluster atoms as immobile, place the $n$ freely mobile (adsorbate) particles randomly outside an exclusion sphere surrounding the cluster, and impose periodic boundary conditions on the simulation cell.

\item \textbf{Iterative contraction:} At iteration $i$ the walker with the highest potential energy, $E_i^{\max}$, is removed and its configuration and energy are recorded. The associated prior-mass increment is
\begin{equation}
w_i = \Gamma_i - \Gamma_{i+1},\qquad
\Gamma_i = \left(\frac{K}{K+1}\right)^{i}.
\label{eq:NS_weights}
\end{equation}

\item \textbf{Replacement:} The discarded walker is replaced by a new configuration generated via a random walk that starts from a randomly cloned surviving walker and is restricted to $E<E_i^{\max}$. 

\item \textbf{Termination:} Iterations continue until the lowest-energy basin is thoroughly sampled.

\end{enumerate}

\subsubsection{Thermodynamic Observables}

The set $\{E(\mathbf{r}_i),w_i\}$ yields the canonical partition function for any inverse temperature $\beta$:
\begin{equation}
Z(N,V,\beta)=\sum_i w_i\,e^{-\beta E(\mathbf{r}_i)},
\label{eq:NS_partition_function}
\end{equation}
where $N$ is the number of particles, $V$ is the system volume, and $E(\mathbf{r}_i)$ is the system's potential energy in configuration $\mathbf{r}_i$ generated at the $i$-th NS iteration.
Ensemble averages follow immediately, for a configuration-dependent observable $A(\mathbf r)$:
\begin{equation}
\langle A\rangle_\beta = \frac{\sum_i w_i\,A(\mathbf r_i)\,e^{-\beta E(\mathbf{r}_i)}}{Z(N, V,\beta)}.
\label{eq:NS_observable_avg}
\end{equation}
In this work, we analyze three key observables. First, we compute the radial distribution function of the adsorbates relative to the cluster center. Second, we determine the average coordination number, $\langle\mathrm{CN}\rangle$, which counts both free atom--free atom and free atom--cluster contacts. Third, we evaluate the constant-volume heat capacity, $C_V(\beta)=k_{\mathrm B}\beta^{2}\,\partial^{2}\ln Z/\partial\beta^{2}$; peaks in $C_V$ reveal coverage-dependent phase transitions.

\subsubsection{Simulation Setup}

Interparticle interactions were described with the LJ pair potential
\begin{equation}
U_\mathrm{LJ} ( r ) = 4 \epsilon \left[ \left( \frac{\sigma}{r} \right)^{12} - \left( \frac{\sigma}{r} \right)^{6} \right],
\label{eq:lj_potential}
\end{equation}
where $r$ is the separation between two atoms, and the parameters $\epsilon$ and $\sigma$ define the energy and length scales, respectively.
A truncated-and-shifted form with a cutoff radius $r_\mathrm{c}=3\sigma$ was used so that $U_{\mathrm{LJ}}(r_\mathrm{c})=0$.

\begin{figure}
\includegraphics[width=0.5\textwidth]{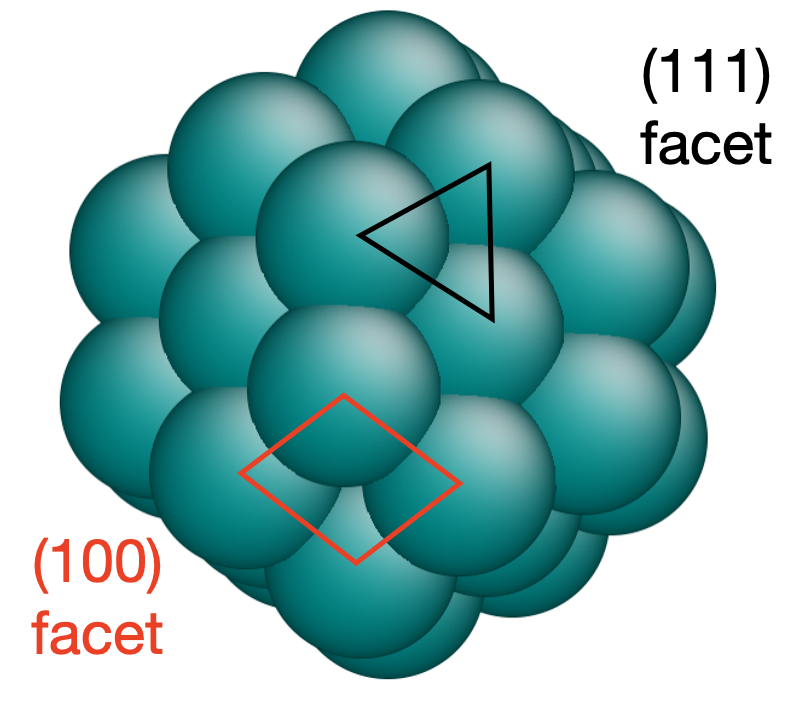}
\caption{\label{fig:lj38}
Cluster of 38 LJ particles in the truncated octahedral global minimum structure, employed as surface in the current work.
The four-coordinated (100) facet and the three-coordinated (111) facet are highlighted by red and black polygons, respectively.
}
\end{figure}

The LJ potential is computationally inexpensive, and its thermodynamic behavior is well characterized, making it a convenient test case for methodological development.
As the model nanocluster, we adopted the 38-particle truncated-octahedral global minimum (LJ$_{38}$, Fig.~\ref{fig:lj38}). \cite{bib:wales_basin_LJ}
Because this structure exposes both (100) and (111) facets, it allows the relative stabilities of adsorbates on these two competing surfaces to be probed within a single simulation.

Three adsorption scenarios were examined (Table~\ref{tab:lj_scenarios}).
Scenario 1 is a reference in which cluster and free atoms share identical LJ parameters: $\sigma_\mathrm{c}=\sigma_\mathrm{f}$ and $\epsilon_\mathrm{c}=\epsilon_\mathrm{f}$ 
 (with LJ parameters $\sigma_\mathrm{c}$ and $\epsilon_\mathrm{c}$ applied for the cluster atoms and $\sigma_\mathrm{f}$ and $\epsilon_\mathrm{f}$ applied for the free moving adsorbate particles).
Scenario 2 reduces the adsorbate well depth to $\epsilon_\mathrm{f} = 0.05\epsilon_\mathrm{c}$ to mimic weak secondary binding. \cite{bichoutskaia2008}
Scenario 3 explores lattice mismatch by scaling the LJ $\sigma$ parameter of the adsorbate to $0.8\sigma_\mathrm{c}$ (negative mismatch, 3a) and $1.2\sigma_\mathrm{c}$ (positive mismatch, 3b).
Mixed interactions follow the Lorentz--Berthelot combining rules in all cases.

\begin{table}
\caption{\label{tab:lj_scenarios}
Lennard--Jones parameters for the three adsorption scenarios.
Subscripts \(\mathrm{c}\) and \(\mathrm{f}\) denote cluster and free atoms, respectively.
Mixed interactions follow the Lorentz--Berthelot rules.
}
\begin{tabular}{lccc}
\hline
Scenario & $\epsilon_{\mathrm{f}}/\epsilon_{\mathrm{c}}$ & $\sigma_{\mathrm{f}}/\sigma_{\mathrm{c}}$ & Purpose / Analogy \\ \hline
1. Equal Interactions & 1.0 & 1.0 & Reference (identical particles) \\
2. Weak Interactions & 0.05 & 1.0 & Weak secondary binding \\
3a. Smaller Adsorbates & 1.0 & 0.8 & Negative lattice mismatch \\
3b. Larger Adsorbates & 1.0 & 1.2 & Positive lattice mismatch \\ \hline
\end{tabular}
\end{table}

The energy-minimized LJ$_{38}$ cluster was placed at the center of a cubic cell of edge length $10\sigma_\mathrm{c}$.
Free particles were inserted at random positions at least $2\sigma_\mathrm{c}$ from the cluster center to avoid large initial overlaps.
During the iterative NS, new configurations were generated by performing a sequence of single-particle Monte Carlo (MC) translation steps on the free particles, under periodic boundary conditions.
Using periodic boundary conditions ensures a finite gas-phase density at the outset without the added complexity of a rigid bounding box; their influence becomes progressively less significant once adsorption dominates, aside from residual finite-size effects.
The MC step size was adjusted at every $K/2$-th iteration to keep the acceptance ratio between 20-50\%.

Sampling parameters were tuned so that independent runs consistently located the global minimum and produced reproducible heat-capacity curves.
For systems containing 5--10 and 11--15 free atoms, we employed $K=2{,}880$ and $K=4{,}032$ walkers, respectively.
When the adsorbate--cluster interaction was weakened (Scenario 2), the number of walkers was increased to $K=8{,}064$ (5--10 adsorbates) and $16{,}128$ (11--15 adsorbates) to adequately sample the larger set of near-degenerate minima.
The random-walk length per NS iteration, $L$, consisted at least $1{,}600$ MC sweeps, where each sweep corresponds to one random attempted translation of every freely moving adsorbate particle.

We have implemented surface nested sampling in the \verb|pymatnest| code and will add a tutorial that reproduces some of our results at \url{https://libatoms.github.io/pymatnest}.

\subsection{Parallel Tempering}

Parallel tempering (PT) surmounts high-energy barriers by allowing replicas at different temperatures to swap configurations, giving it exploration power comparable to NS. \cite{miasojedow2013adaptive,rozada2019effects}
With a well-designed temperature ladder, PT does not, in principle, require any structural bias, although practitioners often start from known low-energy geometries to shorten the equilibration time.
Head-to-head comparisons at constant volume \cite{1st_NS_paper} and constant pressure \cite{pt_phase_dias_ns} show that PT typically needs $\gtrsim$10-fold more energy evaluations than NS to reach the same statistical error in the heat-capacity curve.
This computational cost disparity widens further for first-order transitions with large latent heat. \cite{1st_NS_paper, pt_phase_dias_ns}

Because PT is widely applied to surface phenomena, \cite{terzyk2007hyper,liewehr2016homopolymer,xie2018hamiltonian,mehendale2025effect,ciobanu2004reconstruction} we benchmarked it against NS for the adsorption of 14 LJ adsorbates on the LJ$_{38}$ nanocluster, using identical parameters (Scenario 1).
Our simulations employed the PT implementation in \textsc{LAMMPS} \cite{t2022lammps} with 48 replicas linearly spaced in the reduced temperature range $k_{\mathrm{B}}T/\epsilon_\mathrm{c} = 0.1\!-\!0.8$.
Replica exchanges were attempted every 10,000 molecular-dynamics (MD) steps, and each replica was propagated for $2.75\times10^{7}$ steps.
To gauge sensitivity to the initial guess, we ran three independent sets that began from (i) a random gas-phase configuration, (ii) a local minimum, and (iii) the NS global minimum.
All trajectories used a Langevin thermostat with damping $0.5\,\tau\sqrt{\epsilon_\mathrm{c}/m\sigma_\mathrm{c}^{2}}$.
The data supporting this study's findings are publicly available in reference number \citenum{chatbipho_adsorbate_2025}.

\section{Results and Discussion} \label{sec:results}

We first discuss systems in which the cluster and free particles share equal LJ parameters, detailing their thermodynamic and structural properties as a function of temperature.
Subsequent sections examine two perturbations (scenarios summarized in Table~\ref{tab:lj_scenarios}): (i) same-sized adsorbates that interact more weakly with the surface and (ii) free particles that are either smaller or larger than the substrate particles, introducing lattice mismatch.
For each case, we analyze how these modifications influence phase behavior and the adsorption motifs that emerge.

\subsection{Equal Interactions}\label{sec:equal}

\subsubsection{Phase Transitions}
 
\begin{figure}
\includegraphics[width=\textwidth,angle=90]{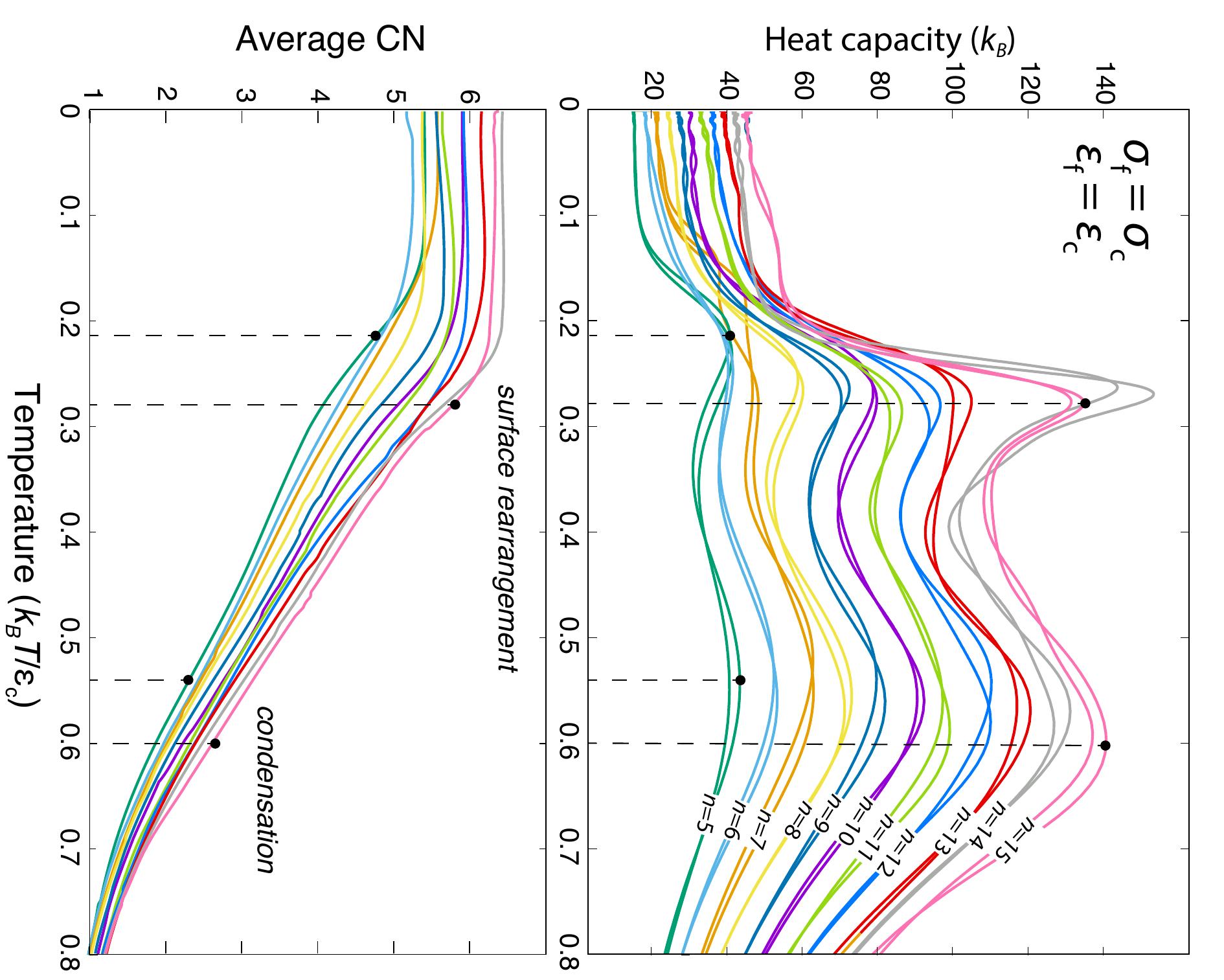}
\caption{\label{fig:LJ38_HC_eps1}
Top panel: Constant volume heat capacity of systems consisting of an LJ$_{38}$ cluster and $n$ free LJ particles, with equal interaction parameters.
Results of two independent nested sampling calculations are shown for each system to demonstrate the level of convergence.
Bottom panel: average coordination number of free particles, as a function of temperature.
Dashed vertical lines mark the position of the heat capacity peaks and corresponding average coordination numbers for one of the $n=5$ and $n=15$ calculations.
}
\end{figure}

From the partition function obtained via nested sampling, we calculated the constant-volume heat capacity as a function of temperature.
Figure~\ref{fig:LJ38_HC_eps1} shows the resulting curves for systems with $n = 5\!-\!15$ free atoms.
Each exhibits two characteristic maxima: a broad high-temperature peak centered at $k_\mathrm{B}T/\epsilon_\mathrm{c} \approx 0.6$ and a lower-temperature, sometimes sharper, peak near $k_\mathrm{B}T/\epsilon_\mathrm{c} \approx 0.25$.

\subsubsection{Coordination Numbers}

To identify the processes responsible for the two heat-capacity maxima, we calculated the average coordination number (CN) of the free particles as a function of temperature, shown in the bottom panel of Fig.~\ref{fig:LJ38_HC_eps1}.
For this, particles were counted to be in the coordination shell if they were closer to each other than $1.35\sigma_\mathrm{c}$, counting only the environments of the free particles. 
At high temperatures, the free particles are in the gas phase, giving low CN ($\lesssim\!1$).
As the system cools, CN increases smoothly and reaches $\sim 2\!-\!3$ around the first $C_V$ peak, signaling condensation of the free particles onto the cluster surface: each free particle now contacts a few cluster particles but has not yet formed ordered surface motifs.
The second $C_V$ peak appears just before CN plateaus; it marks the lateral rearrangement of the condensed layer into configurations that maximize coordination both with the cluster and among the adsorbates themselves.
Overall, the condensation and subsequent surface-ordering transitions mirror those observed previously for flat fcc (100)/(111) surfaces. \cite{yang2024surface}

\subsubsection{Coverage Dependence}

Condensation exhibits a clear, coverage-dependent trend: both the condensation temperature and the height of the corresponding $C_V$ peak increase monotonically with the number of free particles.
The lower-temperature surface-rearrangement peak, however, shows notable coverage-specific variations.
For $n=7$, it is markedly broader than for the other coverages, almost a double peak, implying two closely spaced ordering events.
For $n=14$, the peak sharpens dramatically at $k_\mathrm{B}T/\epsilon_\mathrm{c} \approx 0.28$, potentially signaling a particularly favorable transition, whereas for $n=15$, a subtle shoulder appears at $k_\mathrm{B}T/\epsilon_\mathrm{c} \approx 0.14$.
The structural origin of these features is analyzed in the following sections.

\subsubsection{Facet Ordering}

\begin{figure}
\includegraphics[width=0.9\textwidth,angle=90]{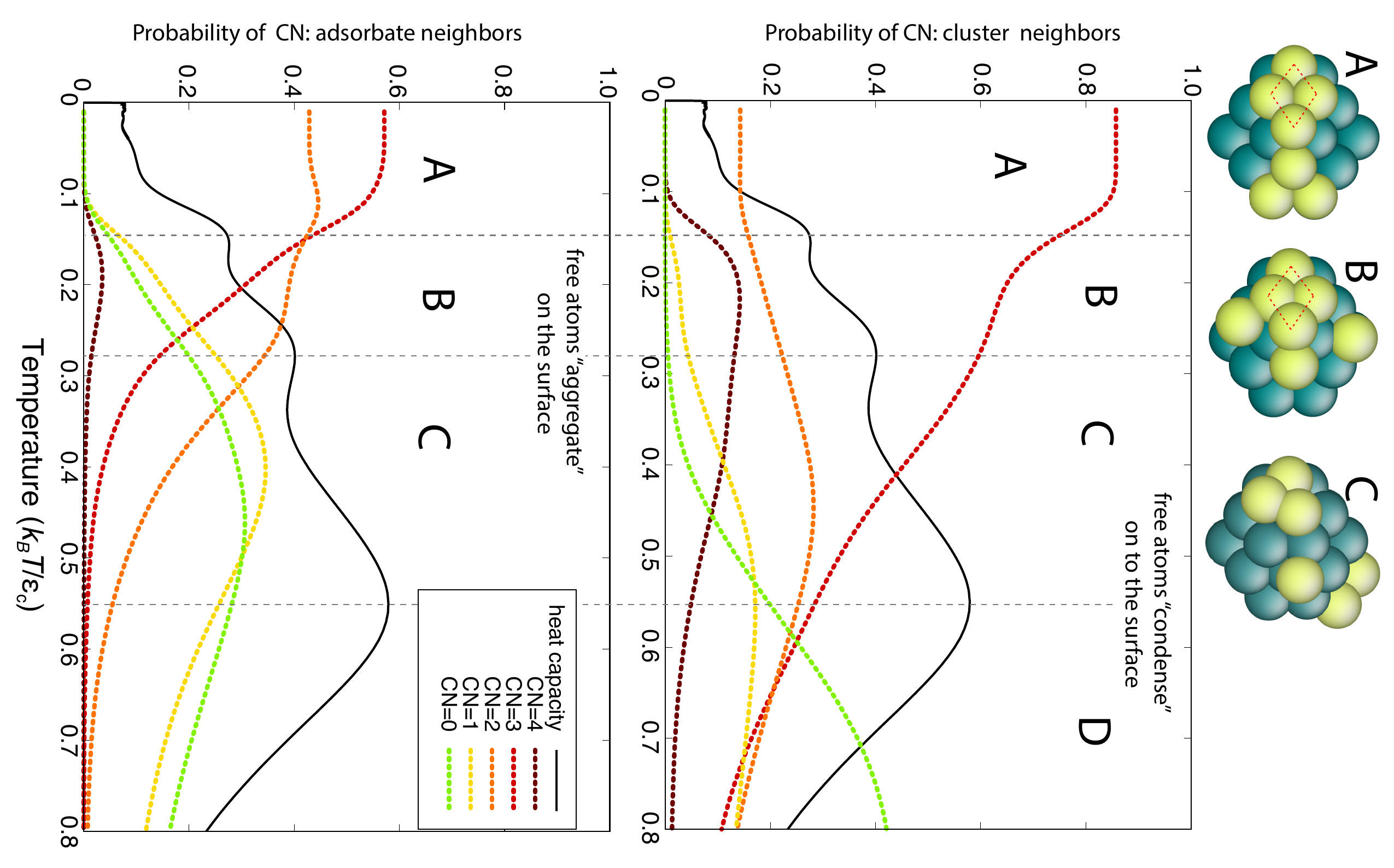}
\caption{\label{fig:LJ38_7adatom_coordination}
Probability of having a specific number of neighbors at a given temperature, in case of the LJ$_{38}$ cluster and seven free LJ particles ($n=7$), with equal interaction parameters.
The top panel shows the coordination taken into account only between free particles and cluster atoms in order to distinguish the occupancy of (111) (threefold-coordinated) and (100) (fourfold-coordinated) sites.
The bottom panel shows the coordination number only between free particles to quantify particle aggregation on the surface.
The heat capacity is shown by a solid black line in both panels, with vertical dashed lines highlighting its peaks.
Snapshots shown above the plots are typical configurations in temperature regions A, B, and C, with the rhombic arrangement of atoms highlighted by red dotted lines.
}
\end{figure}

To characterize the structural transitions associated with the two heat-capacity maxima, we calculated the probability that a free particle has a given number of nearest neighbors.
Because NS provides the phase-space-volume weights, these probabilities can be evaluated as a function of temperature, shown in Figure~\ref{fig:LJ38_7adatom_coordination}.
Figure~\ref{fig:LJ38_7adatom_coordination} separates contacts with cluster particles (top panel) from contacts with other free particles (bottom panel), allowing us to distinguish (i) condensation and facet preference from (ii) lateral aggregation and rearrangement of particles already adsorbed.
For the $n=7$ system, four temperature regions (labeled A--D) emerge.
In Region D ($k_\mathrm{B}T/\epsilon_\mathrm{c}\gtrsim0.55$), a substantial fraction of free particles are in the gas phase, so the probability of having zero neighbors is high.
In Region C, the probability of zero-coordinated particles drops significantly, signaling complete condensation onto the surface.
At the same time, the probabilities of having two or three free-particle neighbors increase, indicating in-plane aggregation, shown in the bottom panel of Fig.~\ref{fig:LJ38_7adatom_coordination}.

In Region B, the occupancies of (111) and (100) sites are essentially fixed (the probabilities of three- and four-fold cluster coordination remain almost constant), yet the lateral arrangement of the condensed layer changes markedly.
The probability of a free particle having three free-particle neighbors triples; visual inspection reveals that four particles frequently form rhombi (red dashed lines on snapshots of typical configurations in the figure), occupying those neighboring hollow sites that correspond to hexagonal close-packed (hcp) stacking.
The remaining three free particles usually adopt one of the four-fold hollow sites on the (100) facets.
In Region A (lowest $T$), the global-minimum structure dominates: five particles occupy hollow sites corresponding to fcc stacking, while the remaining two occupy four-fold hollow sites, with hcp stacking no longer observed.

Thus, the lower-temperature heat-capacity peak reflects a subtle reorganization of an already condensed layer: rhombic hcp motifs give way to the fcc global minimum as the system is cooled.

\subsubsection{Ground States}

\begin{figure}
\includegraphics[width=0.5\textwidth,angle=90]{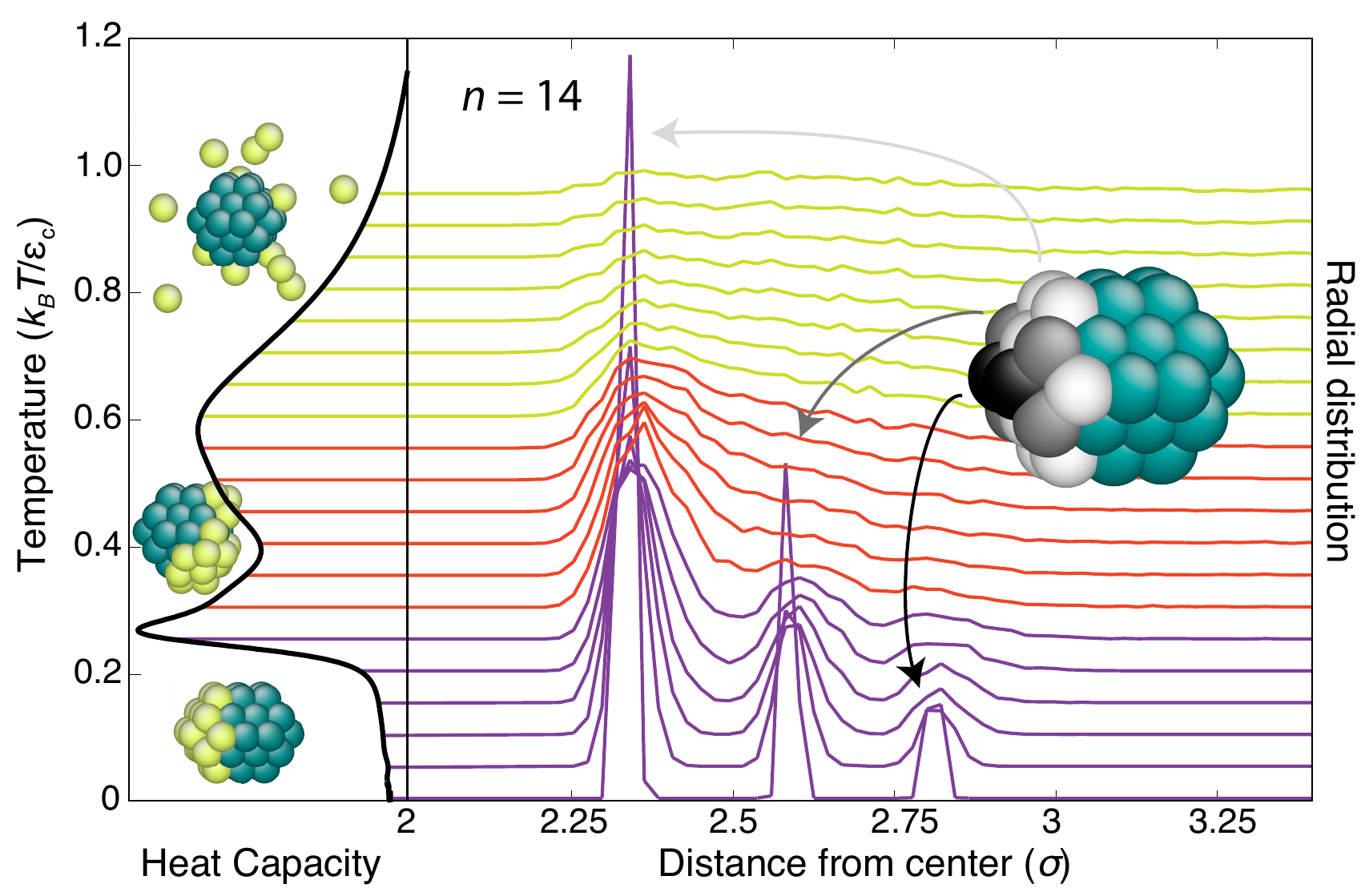}
\caption{\label{fig:LJ38_14_Trdf}
Phase-space-volume-averaged radial distribution of $n=14$ free particles around the center of the cluster.
Lines are shifted for better visibility and for the baselines to correspond to the appropriate temperature on the heat capacity curve, shown in the left-hand panel.
Lines are colored to represent different temperature ranges.
Snapshots of typical structures in each temperature range are shown as illustrations, as well as to demonstrate the distance of various adsorbates from the center of the cluster.
}
\end{figure}

Phase-space-weighted radial distributions of the free particles clarify how the ground states evolve with coverage.
For $n=14$ (shown in Figure~\ref{fig:LJ38_14_Trdf}), the radial distribution is flat above the condensation transition, indicating a gas-like phase.
Cooling first produces a single peak at $r\approx2.3\sigma_\mathrm{c}$ as the free particles adsorb onto the surface, occupying three-fold hollow sites on the (111) facets.
At still lower temperatures, this peak resolves into three sub-peaks, showing that the 14 adsorbates occupy three radially distinct groups of surface sites (hollow sites at $r\approx2.3\sigma_\mathrm{c}$, bridge at $r\approx2.6\sigma_\mathrm{c}$, and atop at $r\approx2.8\sigma_\mathrm{c}$) that together form an fcc ``cap'' on one hemisphere of the cluster.

\begin{figure}
\includegraphics[width=\textwidth]{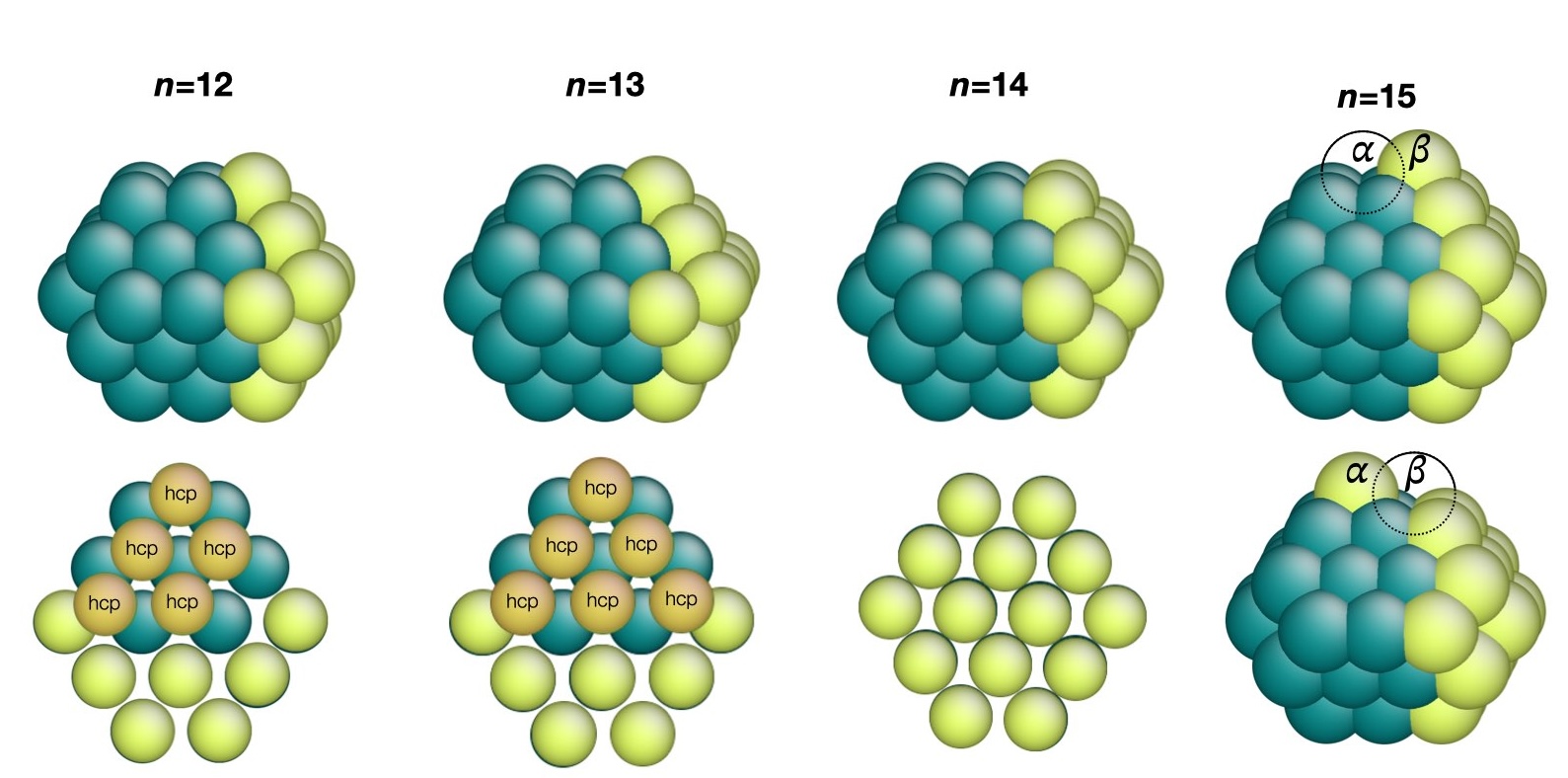}
\caption{\label{fig:LJ12-15_adatoms}
Snapshot of the ground state structures of the LJ$_{38}$ cluster with free particles $n=12-15$.
Atoms corresponding to the fixed cluster are colored dark green, while free atoms are yellow.
The bottom row shows the same structures from a different angle, with atoms occupying hcp positions highlighted in orange.
In the case of $n=15$, the snapshots show two energetically equivalent configurations, with the two four-coordinated surface positions marked by $\alpha$ and $\beta$.
}
\end{figure}

This perfect fcc cap is highly favorable and gives rise to the particularly sharp heat-capacity peak at $k_\mathrm{B}T/\epsilon_\mathrm{c}\approx0.28$, shown in Fig.~\ref{fig:LJ38_HC_eps1}.
For $n<14$, the adsorbates still tend to complete the cap, but one half adopts fcc stacking while the other half adopts hcp stacking, shown in Figure~\ref{fig:LJ12-15_adatoms}.
Adding a $15^\mathrm{th}$ adsorbate leaves the fcc cap intact.
The surplus particle prefers one of the equivalent four-fold hollow sites on a (100) facet (sites $\alpha$ and $\beta$ in Figure~\ref{fig:LJ12-15_adatoms}), where it has four nearest neighbors.
Because these sites are degenerate, the extra free particle is equally likely to occupy any of the equivalent sites in the zero-temperature limit.

\subsubsection{Performance Comparison}

\begin{figure}
\includegraphics[width=0.5\textwidth]{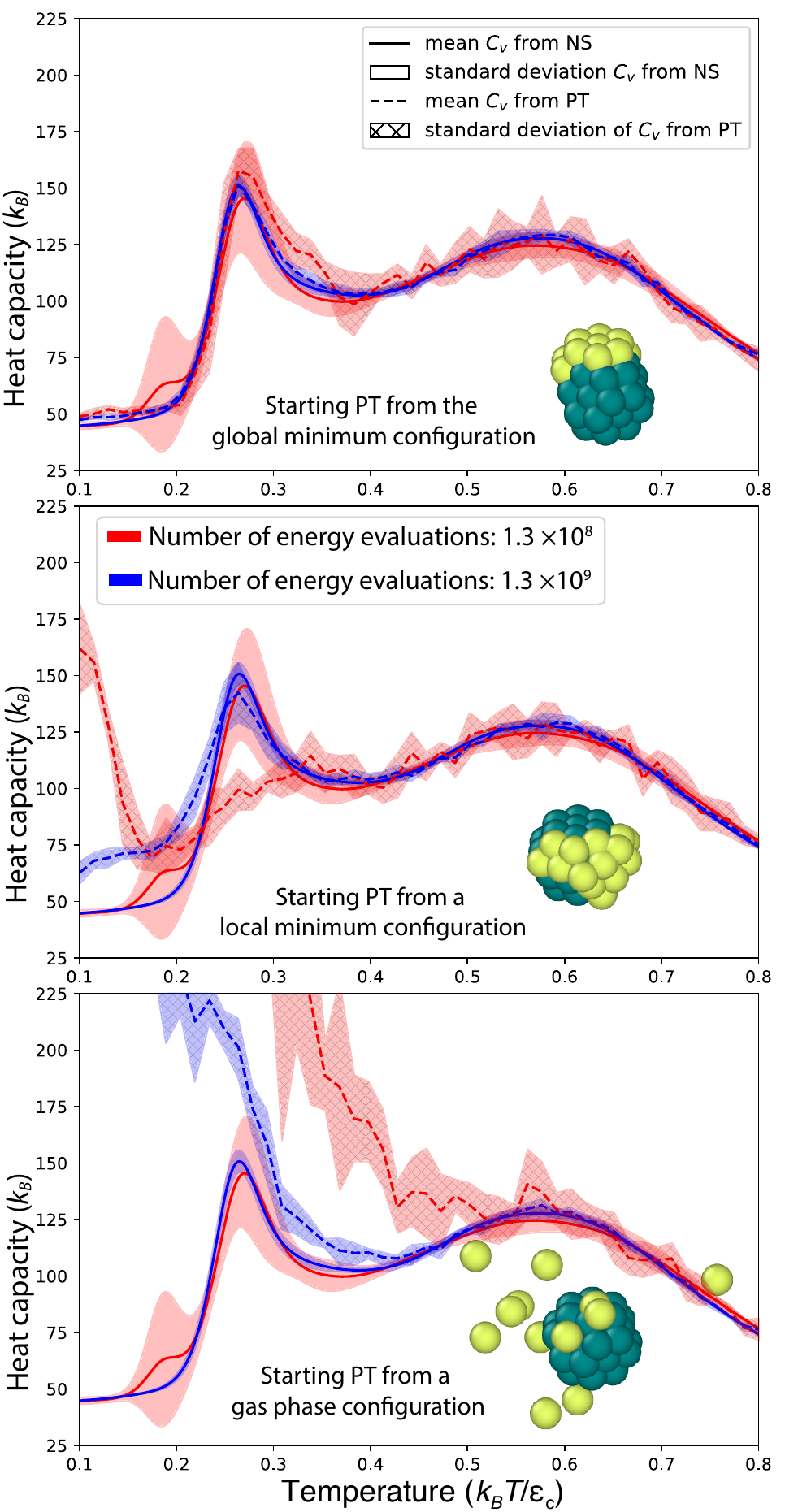}
\caption{\label{fig:NS-PT_compare}
Comparing the computational cost and efficiency of nested sampling (NS) and parallel tempering (PT), by comparing the heat capacities as a function of temperature, for the system of 14 LJ atoms on the surface of the cluster (using the same LJ interaction parameters).
Different panels show different starting configurations (shown as snapshots in the respective panels) for PT: the global minimum structure (top panel), a local minimum structure (middle panel), and a random gas phase configuration (bottom panel).
NS results are identical across panels.
The total number of energy evaluations used to produce the curves is included in the legend.
Shaded areas indicate the standard deviation calculated from five independent runs.
}
\end{figure}

To quantify the cost--accuracy trade-off of NS versus PT, we examined the $n=14$ system, using the constant-volume heat capacity in Figure~\ref{fig:NS-PT_compare}.
Unlike PT, which can be continued until convergence, NS is performed with predetermined $K$ and $L$; combining separate NS runs for higher resolution is non-trivial (because weights cannot be merged \textit{post facto}).
Two sets of NS parameters were compared: the high-resolution setup used earlier in this section (requiring $1.3\times10^9$ energy evaluations for convergence) and a low-resolution setup with fewer walkers and shorter MC walks ($K=576$, $L=600$), reducing the total computational cost by $\sim90$~\% ($1.3\times10^8$ energy evaluations).
The low-resolution NS setup broadens the uncertainty bands, but it still resolves the main phase transitions and locates the global minimum.

The three panels of Fig.~\ref{fig:NS-PT_compare} compare PT runs that start from progressively less informed initial configurations.
When PT is seeded with the NS global minimum, its heat-capacity curve is statistically indistinguishable from that of NS, confirming the latter's accuracy.
Starting PT from a local minimum in which the 14 free particles have already adsorbed onto the cluster surface enables it to resolve the condensation transition within $1.3\times10^8$ energy evaluations, but even a ten-fold increase in computational cost ($1.3\times10^9$ energy evaluations) is insufficient to resolve the lower-temperature surface-rearrangement peak.
NS resolves both transitions at either cost level.
Starting from an unbiased, random gas-phase configuration, PT fails to capture the condensation transition after $1.3\times10^8$ energy evaluations, whereas even the low-resolution NS run resolves both phase transitions and the global minimum; the high-resolution run merely narrows the uncertainty bands.

Overall, NS attains converged thermodynamic observables at equal or lower computational expense than PT unless the true ground state (i.e., the global minimum) is supplied \textit{a priori}.
For realistic problems where that information is unavailable, NS offers the more robust and efficient route to mapping surface phase behavior.

\subsection{Weak Interactions} \label{sec:results-weak}

When cluster--free-particle and free-particle--free-particle interactions are identical, the ground state maximizes each particle's coordination number, so competition between the cluster's facets is minimal.
By contrast, when the free particles interact much more weakly with one another than with the surface ($\epsilon_\mathrm{f}\ll\epsilon_\mathrm{c}$), the four-fold hollow sites on the (100) facets should be energetically favored over the three-fold sites on the (111) facets because they offer stronger binding via higher coordination.
To evaluate how this energetic asymmetry competes with entropic effects, we performed NS calculations with identical $\sigma$ values for all LJ pairs but reduced the adsorbate--adsorbate well depth to $\epsilon_\mathrm{f} = 0.05\epsilon_\mathrm{c}$.

\subsubsection{Phase Transitions}

\begin{figure}
\includegraphics[width=0.5\textwidth,angle=90]{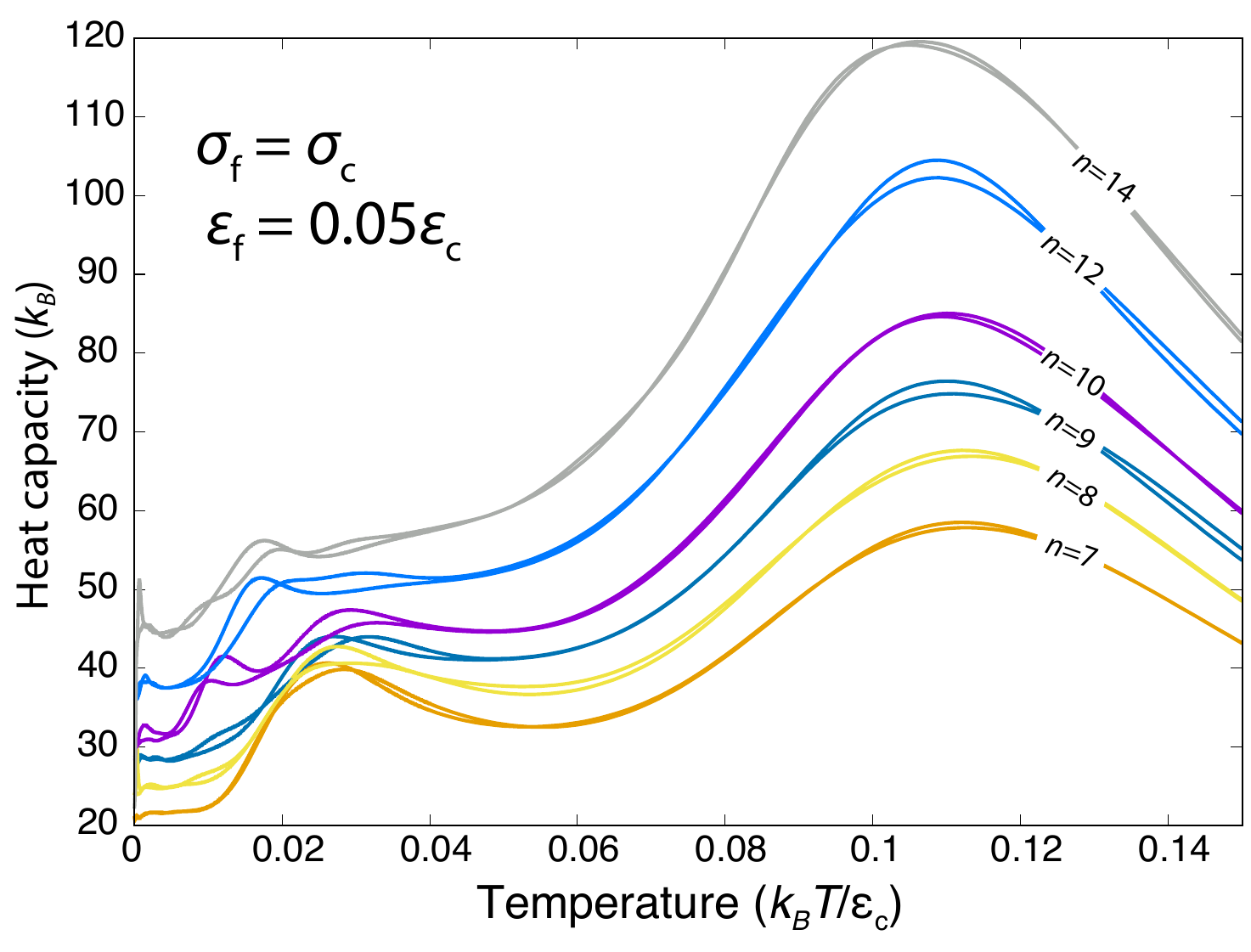}
\caption{\label{fig:LJ38_HC_eps005}
Constant volume heat capacity as a function of temperature, of systems consisting of an LJ$_{38}$ cluster and $n$ free LJ particles, with weaker interaction strength between the free atoms.
}
\end{figure}

Figure~\ref{fig:LJ38_HC_eps005} shows the constant-volume heat capacity for systems in which the adsorbate--adsorbate interaction is weakened to $\epsilon_\mathrm{f}~=~0.05\epsilon_\mathrm{c}$ relative to Scenario 1.
Because all interactions are weaker, both characteristic phase transitions occur at lower reduced temperatures than in the identical-interaction case.
The broad peak at higher temperature $k_\mathrm{B}T/\epsilon_\mathrm{c}\approx0.12$ again signals the condensation of gas-phase free particles onto the cluster surface.
A second, much smaller peak is observed at $k_\mathrm{B}T/\epsilon_\mathrm{c}\approx0.03$; this feature corresponds to surface rearrangement of the adsorbed layer.
Due to the very weak adsorbate--adsorbate interactions, the energy change associated with this surface rearrangement is modest, thus the associated heat-capacity peak is much less pronounced than the condensation peak.

\subsubsection{Facet Ordering}

Because free particles interact much more strongly with cluster atoms than with one another when $\epsilon_\mathrm{f}\ll\epsilon_\mathrm{c}$, the four-fold hollow sites on the (100) facets offer stronger binding.
All six (100) sites are occupied first, whereas three-fold hollow sites on the (111) facets remain vacant until $n>6$.
This energetic preference produces additional low-temperature features in the heat-capacity curves, but the underlying energy differences are only a few times $10^{-3}\epsilon_\mathrm{c}$; resolving such small energy differences makes full convergence challenging, so we focus on the representative $n=14$ system.

\begin{figure}
\includegraphics[width=\textwidth,angle=90]{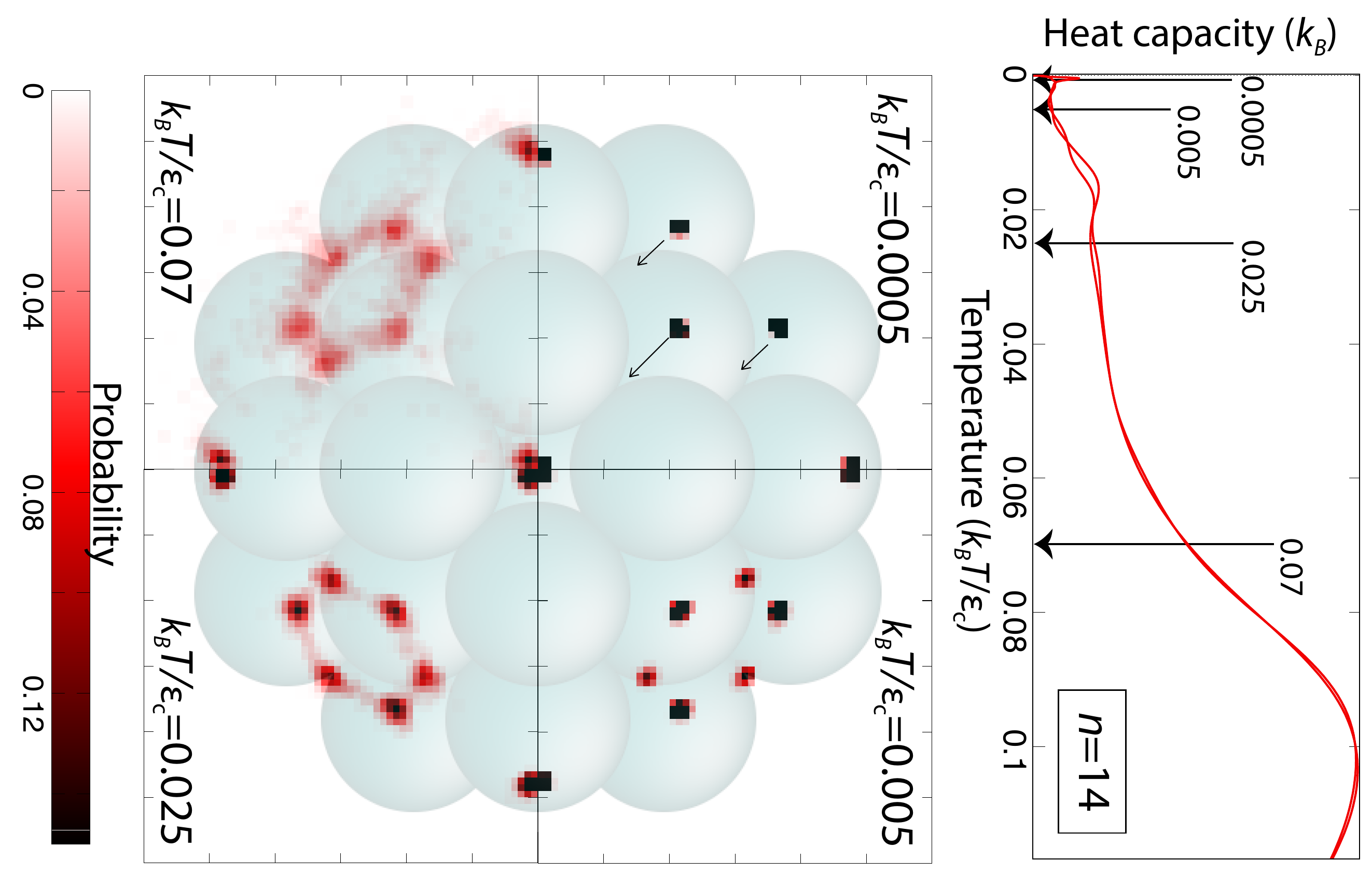}
\caption{\label{fig:LJ38_14adatom}
$n=14$ free particles on the surface of LJ$_{38}$ cluster, with weak free particle interactions.
Top panel: heat capacity curve, with arrows pointing to temperatures where the weighted average free particle distribution has been calculated.
Bottom panel: weighted average of positions of free particles on the surface.
The light green structure shows the LJ$_{38}$ configuration, with red points demonstrating the probability of finding free particles at a particular position at four different temperatures.
Each quadrant of the panel shows results averaged over all quadrants of the cluster.
Arrows on the top left segment highlight hollow sites corresponding to hcp positions.
}
\end{figure}

Figure~\ref{fig:LJ38_14adatom} links the heat-capacity curve of the $n=14$ system to the probability of finding the adsorbate particles in certain locations, for four characteristic temperatures.
At $k_\mathrm{B}T/\epsilon_\mathrm{c}=0.0005$ (basin of attraction of the global minimum), the free particles occupy all six (100) hollow sites and the edge-adjacent hcp-stacked hollow sites on the (111) facets.
At $k_\mathrm{B}T/ \epsilon_\mathrm{c}=0.005$, local minima becomes available with the free particles not only occupy the same sites as at $k_\mathrm{B}T/\epsilon_\mathrm{c}=0.0005$, but also partially occupy the fcc-stacked hollow sites adjacent to the hcp-stacked ones.
At $k_\mathrm{B}T/ \epsilon_\mathrm{c}=0.025$, fcc and hcp sites are now roughly equally populated, and the probability of finding a free particle on the (111) facets spreads evenly, indicating increased lateral mobility of three-fold-coordinated free particles while the four-fold-coordinated free particles remain pinned.
At $k_\mathrm{B}T/ \epsilon_\mathrm{c}=0.07$, a broad ensemble of surface configurations is sampled.

\subsubsection{Surface Diffusion}

To test whether the characteristic surface arrangements are local minima and to compare adsorbate mobility across sites, we ran $NVT$ MD simulations, starting from configurations generated by NS.
At $k_\mathrm{B}T/\epsilon_\mathrm{c}=0.07$, the free particles are mobile, with three-fold-coordinated adsorbates on the (111) facets diffusing most readily.
Diffusivity falls sharply on cooling---by $k_\mathrm{B}T/\epsilon_\mathrm{c}=0.025$, only particles on partially filled (111) facets hop between hollow sites, and no inter-facet motion is observed.
The inter-facet barrier is therefore already prohibitive, yielding a rugged landscape of nearly degenerate minima.
This frustration highlights nested sampling's top-down exploration, which captures the subtle competition between facets and the resulting local arrangements without structural bias.

\subsection{Size-Mismatch Effects}

Changing the LJ $\sigma$ parameter, and thus the effective size of the adsorbing particles, has a significant impact on adsorption.
Although the overall phase sequence remains unchanged (gas-phase condensation at higher temperature followed by surface restructuring at lower temperature), the magnitude and character of each transition depend sensitively on $\sigma$.

\subsubsection{Phase Transitions}

\begin{figure}
\includegraphics[width=0.5\textwidth]{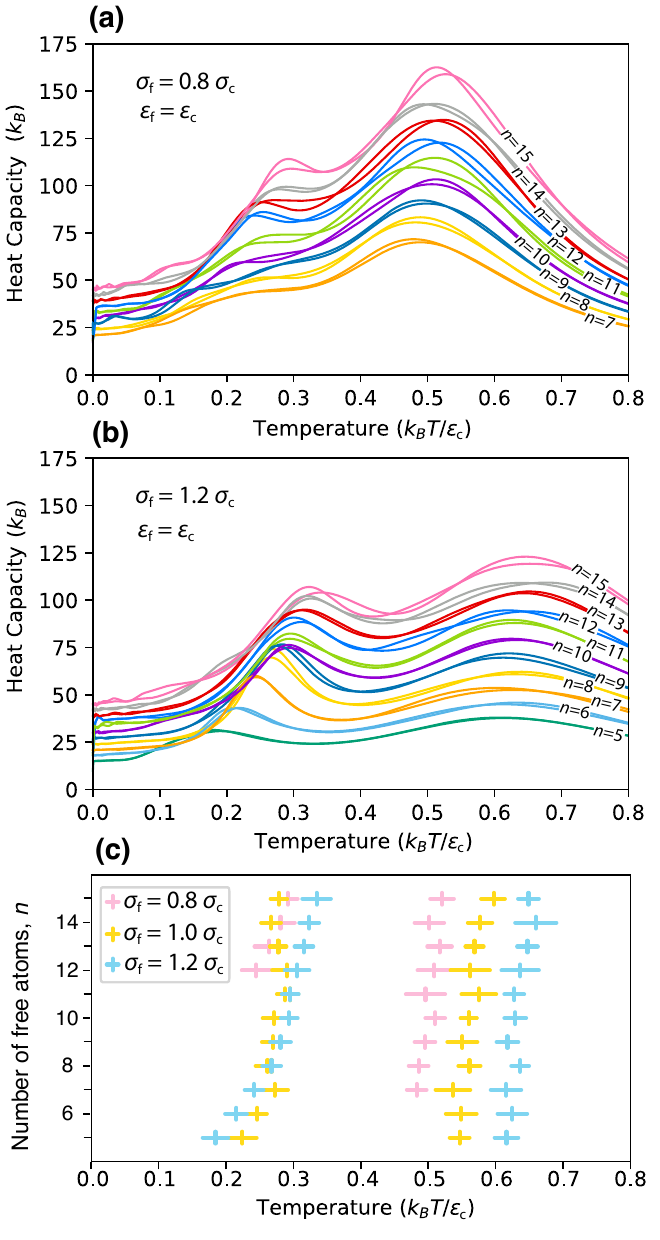}
\caption{\label{fig:cv_sigma}
Constant volume heat capacity as a function of temperature, of systems consisting of an LJ$_{38}$ cluster and $n$ free LJ particles, with the free atoms having smaller effective size (panel (a)) and larger effective size (panel (b)).
Panel (c) compares the location of heat capacity peaks at different $\sigma_f$ values.
The crosses indicate the average transition temperature obtained from two independent NS runs, while the horizontal bars represent the range between these two values.
}
\end{figure}

Figure~\ref{fig:cv_sigma} shows constant-volume heat-capacity curves for systems in which the adsorbate diameter is either 20\% smaller ($\sigma_\mathrm{f}=0.8\sigma_\mathrm{c}$, Scenario 3a) or 20\% larger ($\sigma_\mathrm{f}=1.2\sigma_\mathrm{c}$, Scenario 3b) than that of the cluster particles.
Increasing $\sigma_\mathrm{f}$ from $0.8\sigma_\mathrm{c}$ to $1.2\sigma_\mathrm{c}$ shifts the condensation peak to slightly higher reduced temperatures and broadens it.
This broadening is consistent with the longer-range cluster--adsorbate attraction and with geometric frustration that inhibits dense, commensurate packing of the larger adsorbates.
The lower-temperature peak associated with lateral surface rearrangement is less pronounced for either size mismatch and almost absent when $\sigma_\mathrm{f}=0.8\sigma_\mathrm{c}$ with $n<11$, because the smaller adsorbates interact over a shorter range and, at these low coverages, their mutual contacts are too sparse to compete with the stronger adsorbate--cluster binding.
Only when $\sigma_\mathrm{f}~=~\sigma_\mathrm{c}$ do snapshots reveal a commensurate close-packed overlayer (see Section~\ref{sec:equal}).
Any size mismatch introduces incommensurability, frustrating this cooperative rearrangement and suppressing its heat-capacity signature.
Panel (c) of Figure~\ref{fig:cv_sigma} summarizes the effect of the LJ $\sigma$ parameter on the phase-transition temperatures.

\subsubsection{Coordination Numbers} \label{sec:mismatch-cns}

\begin{figure}
\includegraphics[width=\textwidth]{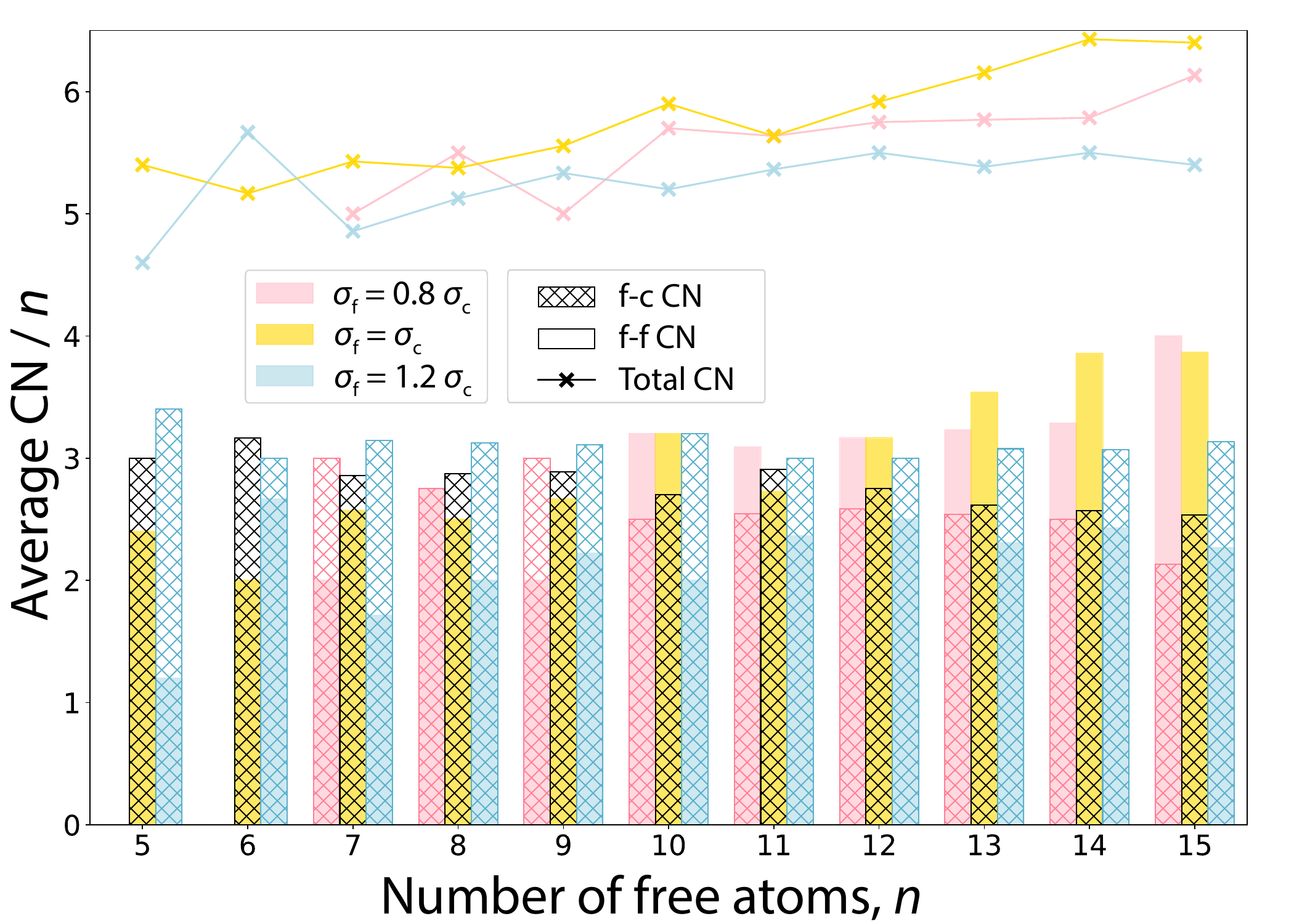}
\caption{\label{fig:CN_sigma}
Average coordination number of free particles in the global minimum structure, as a function of the system's size.
Bar charts show the coordination number, taking into account only free particle--free particle neighbors (solid colors) and taking into account only free particle--cluster-atom neighbors (cross-hatched colors).
Lines and symbols show the total coordination number.
Different colors represent different scenarios of LJ $\sigma$ parameters.
}
\end{figure}

Figure~\ref{fig:CN_sigma} decomposes the average CN per adsorbate into adsorbate--cluster (f--c) and adsorbate--adsorbate (f--f) contributions for every coverage and size-mismatch scenario.
This decomposition highlights how lattice mismatch reshapes the driving forces for adsorption and subsequent surface rearrangement.

\begin{figure}
\includegraphics[width=\textwidth]{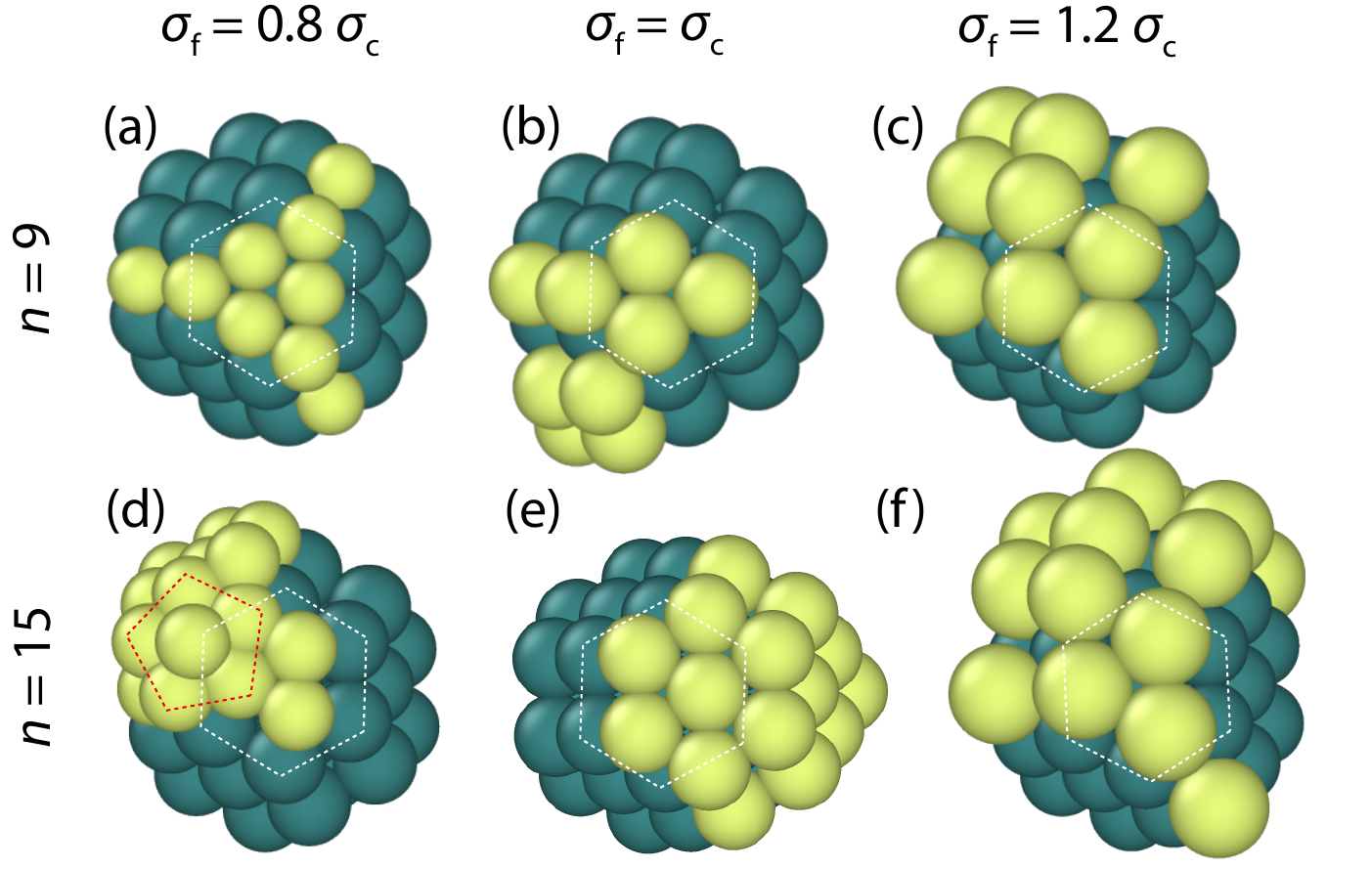}
\caption{\label{fig:9adatom_dif_sigma}
Snapshots of the global minimum structures identified by nested sampling, using $n=9$ (top row) and $n=15$ (bottom row) free particles (shown by yellow), and different interaction parameters: $\sigma_\mathrm{f} = 0.8\sigma_\mathrm{c}$ (panels (a) and (d)), $\sigma_\mathrm{f} = \sigma_\mathrm{c}$ (panels (b) and (e)) and $\sigma_\mathrm{f} = 1.2\sigma_\mathrm{c}$ (panels (c) and (f)).
White dashed hexagons highlight one of the (111) facets, and the red dashed polygon on panel (d) highlights the pentagonal arrangement of adsorbates.
}
\end{figure}

For smaller free particles ($\sigma_\mathrm{f} = 0.8\sigma_\mathrm{c}$), the three-fold hollow sites on the (111) facets are the first sites occupied.
Once a (111) facet is saturated, the hollow--hollow spacing keeps the smaller free particles farther apart than their equilibrium f--f distance, so additional free particles adsorb at edges or vertices, whose less-constrained geometry lets adsorbates slide laterally and approach one another more closely, thereby maximizing f--f contacts.
The $n=15$ global minimum is a notable example: five adsorbates form an approximately regular pentagonal ring capped by a central particle with no cluster-atom neighbors (as shown in Fig.~\ref{fig:9adatom_dif_sigma} panel (d)), an icosahedral motif reminiscent of LJ clusters. \cite{bib:wales_basin_LJ}

%Conversely, larger free particles ($\sigma_\mathrm{f} = 1.2\sigma_\mathrm{c}$) would crowd each other if they occupied adjacent three-fold hollow sites on the (111) facets.
Conversely, the energy minimum distance between larger free particles ($\sigma_\mathrm{f} = 1.2\sigma_\mathrm{c}$) is longer than the distance between adjacent three-fold hollow sites, thus adsorbate particles would repel each other if saturated a (111) facet.
To avoid crowding, the first adsorbates bind in the isolated four-fold hollow sites on the (100) facets, where strong adsorbate--cluster coordination outweighs weak lateral bonding.
Figure~\ref{fig:CN_sigma} confirms this interpretation: the coordination number between free particles and cluster particles is consistently higher than when only free-particle neighbors are counted, and the free-particle--free-particle coordination is considerably lower than for the other size ratios.

Because each size ratio stabilizes a different low-energy motif, the total CN (line plots in Figure~\ref{fig:CN_sigma}) no longer grows smoothly with $n$ when $\sigma_\mathrm{f} \neq \sigma_\mathrm{c}$.
Instead, it plateaus or even decreases whenever a new layer or an incommensurate overlayer becomes favorable.
Representative ground-state structures for $n=9$ and $n=15$ (Figure~\ref{fig:9adatom_dif_sigma}) illustrate these competing motifs.

\subsubsection{Temperature Dependence}

\begin{figure}
\includegraphics[width=0.5\textwidth]{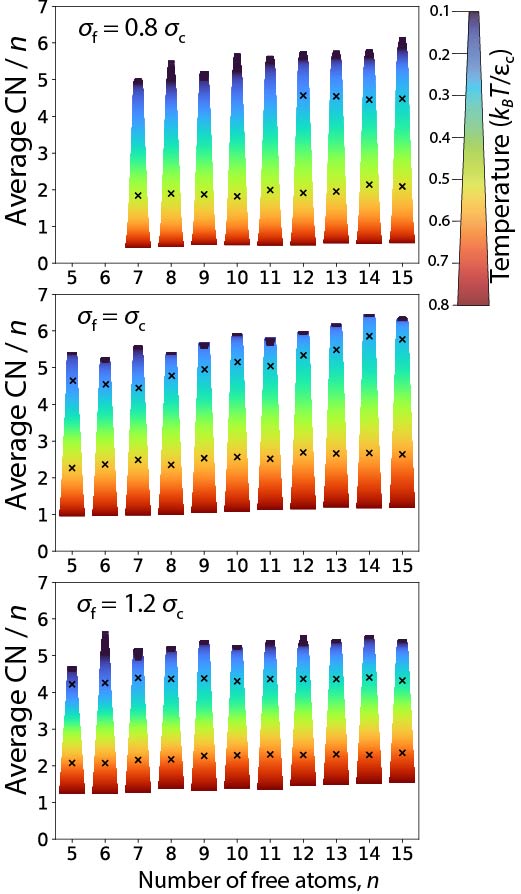}
\caption{\label{fig:CN_T_all}
Average total coordination number of adsorbate particles for different system sizes.
Color scale and bar width represent the temperatures, with black crosses indicating the average transition temperatures obtained from two independent NS runs.
}
\end{figure}

The temperature dependence of CN (Figure~\ref{fig:CN_T_all}) reinforces the picture established in Section~\ref{sec:mismatch-cns}.
When $\sigma_\mathrm{f} = \sigma_\mathrm{c}$, CN minimally changes below the low-temperature transition (i.e., above the black crosses at higher CNs in Figure~\ref{fig:CN_T_all}), indicating only limited rearrangement among similar minima.
For larger free particles, the transition CN is nearly coverage-independent, reflecting the limited palette of under-coordinated motifs imposed by steric mismatch.
Smaller free particles, by contrast, exhibit large, irregular jumps in CN at the transition, which is evidence of significant surface rearrangement and multiple, nearly degenerate adsorption motifs persisting to low temperatures.

Together, these trends show that lattice mismatch shifts the balance between adsorbate--surface and adsorbate--adsorbate interactions, dictating whether layer completion, incommensurability, or multiple-layer formation dominates adsorption thermodynamics.

\section{Conclusions}

We applied surface nested sampling (NS) \cite{yang2024surface} to a Lennard--Jones (LJ) truncated-octahedral nanocluster (LJ$_{38}$) hosting freely mobile adsorbates.
By varying surface coverage, interaction strength, and size mismatch, we computed the canonical partition function, calculated heat-capacity curves, and analyzed coordination statistics across a broad temperature range.
Two generic, coverage-dependent transitions emerged: condensation at higher temperatures and surface reordering at lower temperatures.
Their character shifts systematically with interaction parameters and available adsorption sites.
Uniform interactions favor formation of an ordered overlayer; weakened adsorbate--adsorbate interactions drive preferential occupation of four-fold sites on (100) facets; size mismatch changes the dominant driving force, smaller adsorbates maximize adsorbate--adsorbate contacts, whereas larger ones maximize adsorbate--surface contacts, often resulting in under-coordination.
Benchmarking against parallel tempering (PT) shows that, unless a pre-identified global minimum is supplied a priori, NS achieves converged thermodynamic properties at consistently lower computational cost while remaining free of structural bias.
These results demonstrate that NS offers an efficient, unbiased route to adsorption thermodynamics and phase behavior on complex surfaces.
Extending the framework to realistic interaction potentials, heterogeneous surfaces, and multi-component adsorbates should further broaden its utility in surface science.

\begin{acknowledgments}

LBP acknowledges support from the EPSRC through an individual Early Career Fellowship (EP/T000163/1).
RBW acknowledges support from the National Science Foundation under Grant No.~2305155.
TC acknowledges PhD sponsorship from the Institute for the Promotion of Teaching Science and Technology, Thailand.
LBP and TC used computing facilities provided by the Scientific Computing Research Technology Platform of the University of Warwick.
A portion of this research was conducted at the Center for Nanophase Materials Sciences, which is a DOE Office of Science User Facility.

\end{acknowledgments}

\section*{Data Availability}

The data that support the findings of this study are openly available in Warwick Research Archive Portal at https://doi.org/[doi], reference number \citenum{chatbipho_adsorbate_2025}.

\bibliography{ns-paper.bib}

\begin{thebibliography}{59}%
\makeatletter
\providecommand \@ifxundefined [1]{%
 \@ifx{#1\undefined}
}%
\providecommand \@ifnum [1]{%
 \ifnum #1\expandafter \@firstoftwo
 \else \expandafter \@secondoftwo
 \fi
}%
\providecommand \@ifx [1]{%
 \ifx #1\expandafter \@firstoftwo
 \else \expandafter \@secondoftwo
 \fi
}%
\providecommand \natexlab [1]{#1}%
\providecommand \enquote  [1]{``#1''}%
\providecommand \bibnamefont  [1]{#1}%
\providecommand \bibfnamefont [1]{#1}%
\providecommand \citenamefont [1]{#1}%
\providecommand \href@noop [0]{\@secondoftwo}%
\providecommand \href [0]{\begingroup \@sanitize@url \@href}%
\providecommand \@href[1]{\@@startlink{#1}\@@href}%
\providecommand \@@href[1]{\endgroup#1\@@endlink}%
\providecommand \@sanitize@url [0]{\catcode `\\12\catcode `\$12\catcode `\&12\catcode `\#12\catcode `\^12\catcode `\_12\catcode `\%12\relax}%
\providecommand \@@startlink[1]{}%
\providecommand \@@endlink[0]{}%
\providecommand \url  [0]{\begingroup\@sanitize@url \@url }%
\providecommand \@url [1]{\endgroup\@href {#1}{\urlprefix }}%
\providecommand \urlprefix  [0]{URL }%
\providecommand \Eprint [0]{\href }%
\providecommand \doibase [0]{http://dx.doi.org/}%
\providecommand \selectlanguage [0]{\@gobble}%
\providecommand \bibinfo  [0]{\@secondoftwo}%
\providecommand \bibfield  [0]{\@secondoftwo}%
\providecommand \translation [1]{[#1]}%
\providecommand \BibitemOpen [0]{}%
\providecommand \bibitemStop [0]{}%
\providecommand \bibitemNoStop [0]{.\EOS\space}%
\providecommand \EOS [0]{\spacefactor3000\relax}%
\providecommand \BibitemShut  [1]{\csname bibitem#1\endcsname}%
\let\auto@bib@innerbib\@empty
%</preamble>
\bibitem [{\citenamefont {Pomerantseva}\ \emph {et~al.}(2019)\citenamefont {Pomerantseva}, \citenamefont {Bonaccorso}, \citenamefont {Feng}, \citenamefont {Cui},\ and\ \citenamefont {Gogotsi}}]{pomerantseva2019energy}%
  \BibitemOpen
  \bibfield  {author} {\bibinfo {author} {\bibfnamefont {E.}~\bibnamefont {Pomerantseva}}, \bibinfo {author} {\bibfnamefont {F.}~\bibnamefont {Bonaccorso}}, \bibinfo {author} {\bibfnamefont {X.}~\bibnamefont {Feng}}, \bibinfo {author} {\bibfnamefont {Y.}~\bibnamefont {Cui}}, \ and\ \bibinfo {author} {\bibfnamefont {Y.}~\bibnamefont {Gogotsi}},\ }\href@noop {} {\bibfield  {journal} {\bibinfo  {journal} {Science}\ }\textbf {\bibinfo {volume} {366}},\ \bibinfo {pages} {eaan8285} (\bibinfo {year} {2019})}\BibitemShut {NoStop}%
\bibitem [{\citenamefont {Buseck}\ and\ \citenamefont {Adachi}(2008)}]{buseck2008nanoparticles}%
  \BibitemOpen
  \bibfield  {author} {\bibinfo {author} {\bibfnamefont {P.~R.}\ \bibnamefont {Buseck}}\ and\ \bibinfo {author} {\bibfnamefont {K.}~\bibnamefont {Adachi}},\ }\href@noop {} {\bibfield  {journal} {\bibinfo  {journal} {Elements}\ }\textbf {\bibinfo {volume} {4}},\ \bibinfo {pages} {389} (\bibinfo {year} {2008})}\BibitemShut {NoStop}%
\bibitem [{\citenamefont {Rabajczyk}\ \emph {et~al.}(2020)\citenamefont {Rabajczyk}, \citenamefont {Zielecka}, \citenamefont {Porowski},\ and\ \citenamefont {Hopke}}]{rabajczyk2020metal}%
  \BibitemOpen
  \bibfield  {author} {\bibinfo {author} {\bibfnamefont {A.}~\bibnamefont {Rabajczyk}}, \bibinfo {author} {\bibfnamefont {M.}~\bibnamefont {Zielecka}}, \bibinfo {author} {\bibfnamefont {R.}~\bibnamefont {Porowski}}, \ and\ \bibinfo {author} {\bibfnamefont {P.~K.}\ \bibnamefont {Hopke}},\ }\href@noop {} {\bibfield  {journal} {\bibinfo  {journal} {Environmental Science: Nano}\ }\textbf {\bibinfo {volume} {7}},\ \bibinfo {pages} {3233} (\bibinfo {year} {2020})}\BibitemShut {NoStop}%
\bibitem [{\citenamefont {Liu}\ and\ \citenamefont {Corma}(2018)}]{liu2018metal}%
  \BibitemOpen
  \bibfield  {author} {\bibinfo {author} {\bibfnamefont {L.}~\bibnamefont {Liu}}\ and\ \bibinfo {author} {\bibfnamefont {A.}~\bibnamefont {Corma}},\ }\href@noop {} {\bibfield  {journal} {\bibinfo  {journal} {Chemical reviews}\ }\textbf {\bibinfo {volume} {118}},\ \bibinfo {pages} {4981} (\bibinfo {year} {2018})}\BibitemShut {NoStop}%
\bibitem [{\citenamefont {Salata}(2004)}]{bio-nano}%
  \BibitemOpen
  \bibfield  {author} {\bibinfo {author} {\bibfnamefont {O.~V.}\ \bibnamefont {Salata}},\ }\href@noop {} {\bibfield  {journal} {\bibinfo  {journal} {Journal of nanobiotechnology}\ }\textbf {\bibinfo {volume} {2}},\ \bibinfo {pages} {1} (\bibinfo {year} {2004})}\BibitemShut {NoStop}%
\bibitem [{\citenamefont {Khan}\ \emph {et~al.}(2019)\citenamefont {Khan}, \citenamefont {Saeed},\ and\ \citenamefont {Khan}}]{bio-nano2}%
  \BibitemOpen
  \bibfield  {author} {\bibinfo {author} {\bibfnamefont {I.}~\bibnamefont {Khan}}, \bibinfo {author} {\bibfnamefont {K.}~\bibnamefont {Saeed}}, \ and\ \bibinfo {author} {\bibfnamefont {I.}~\bibnamefont {Khan}},\ }\href@noop {} {\bibfield  {journal} {\bibinfo  {journal} {Arabian journal of chemistry}\ }\textbf {\bibinfo {volume} {12}},\ \bibinfo {pages} {908} (\bibinfo {year} {2019})}\BibitemShut {NoStop}%
\bibitem [{\citenamefont {Mitchell}\ \emph {et~al.}(2021)\citenamefont {Mitchell}, \citenamefont {Billingsley}, \citenamefont {Haley}, \citenamefont {Wechsler}, \citenamefont {Peppas},\ and\ \citenamefont {Langer}}]{mitchell2021engineering}%
  \BibitemOpen
  \bibfield  {author} {\bibinfo {author} {\bibfnamefont {M.~J.}\ \bibnamefont {Mitchell}}, \bibinfo {author} {\bibfnamefont {M.~M.}\ \bibnamefont {Billingsley}}, \bibinfo {author} {\bibfnamefont {R.~M.}\ \bibnamefont {Haley}}, \bibinfo {author} {\bibfnamefont {M.~E.}\ \bibnamefont {Wechsler}}, \bibinfo {author} {\bibfnamefont {N.~A.}\ \bibnamefont {Peppas}}, \ and\ \bibinfo {author} {\bibfnamefont {R.}~\bibnamefont {Langer}},\ }\href@noop {} {\bibfield  {journal} {\bibinfo  {journal} {Nature reviews drug discovery}\ }\textbf {\bibinfo {volume} {20}},\ \bibinfo {pages} {101} (\bibinfo {year} {2021})}\BibitemShut {NoStop}%
\bibitem [{\citenamefont {Haruta}\ \emph {et~al.}(1987)\citenamefont {Haruta}, \citenamefont {Kobayashi}, \citenamefont {Sano},\ and\ \citenamefont {Yamada}}]{haruta1987novel}%
  \BibitemOpen
  \bibfield  {author} {\bibinfo {author} {\bibfnamefont {M.}~\bibnamefont {Haruta}}, \bibinfo {author} {\bibfnamefont {T.}~\bibnamefont {Kobayashi}}, \bibinfo {author} {\bibfnamefont {H.}~\bibnamefont {Sano}}, \ and\ \bibinfo {author} {\bibfnamefont {N.}~\bibnamefont {Yamada}},\ }\href@noop {} {\bibfield  {journal} {\bibinfo  {journal} {Chemistry Letters}\ }\textbf {\bibinfo {volume} {16}},\ \bibinfo {pages} {405} (\bibinfo {year} {1987})}\BibitemShut {NoStop}%
\bibitem [{\citenamefont {Gupta}\ \emph {et~al.}(2007)\citenamefont {Gupta}, \citenamefont {Naregalkar}, \citenamefont {Vaidya},\ and\ \citenamefont {Gupta}}]{NP_medicine}%
  \BibitemOpen
  \bibfield  {author} {\bibinfo {author} {\bibfnamefont {A.~K.}\ \bibnamefont {Gupta}}, \bibinfo {author} {\bibfnamefont {R.~R.}\ \bibnamefont {Naregalkar}}, \bibinfo {author} {\bibfnamefont {V.~D.}\ \bibnamefont {Vaidya}}, \ and\ \bibinfo {author} {\bibfnamefont {M.}~\bibnamefont {Gupta}},\ }\href@noop {} {\bibfield  {journal} {\bibinfo  {journal} {Nanomedicine}\ }\textbf {\bibinfo {volume} {2}},\ \bibinfo {pages} {23} (\bibinfo {year} {2007})}\BibitemShut {NoStop}%
\bibitem [{\citenamefont {Issa}\ \emph {et~al.}(2013)\citenamefont {Issa}, \citenamefont {Obaidat}, \citenamefont {Albiss},\ and\ \citenamefont {Haik}}]{NP_magnetic}%
  \BibitemOpen
  \bibfield  {author} {\bibinfo {author} {\bibfnamefont {B.}~\bibnamefont {Issa}}, \bibinfo {author} {\bibfnamefont {I.~M.}\ \bibnamefont {Obaidat}}, \bibinfo {author} {\bibfnamefont {B.~A.}\ \bibnamefont {Albiss}}, \ and\ \bibinfo {author} {\bibfnamefont {Y.}~\bibnamefont {Haik}},\ }\href@noop {} {\bibfield  {journal} {\bibinfo  {journal} {International journal of molecular sciences}\ }\textbf {\bibinfo {volume} {14}},\ \bibinfo {pages} {21266} (\bibinfo {year} {2013})}\BibitemShut {NoStop}%
\bibitem [{\citenamefont {Stark}\ \emph {et~al.}(2015)\citenamefont {Stark}, \citenamefont {Stoessel}, \citenamefont {Wohlleben},\ and\ \citenamefont {Hafner}}]{NP_industrial}%
  \BibitemOpen
  \bibfield  {author} {\bibinfo {author} {\bibfnamefont {W.~J.}\ \bibnamefont {Stark}}, \bibinfo {author} {\bibfnamefont {P.~R.}\ \bibnamefont {Stoessel}}, \bibinfo {author} {\bibfnamefont {W.}~\bibnamefont {Wohlleben}}, \ and\ \bibinfo {author} {\bibfnamefont {A.}~\bibnamefont {Hafner}},\ }\href@noop {} {\bibfield  {journal} {\bibinfo  {journal} {Chemical Society Reviews}\ }\textbf {\bibinfo {volume} {44}},\ \bibinfo {pages} {5793} (\bibinfo {year} {2015})}\BibitemShut {NoStop}%
\bibitem [{\citenamefont {Afolalu}\ \emph {et~al.}(2019)\citenamefont {Afolalu}, \citenamefont {Soetan}, \citenamefont {Ongbali}, \citenamefont {Abioye},\ and\ \citenamefont {Oni}}]{NP_review}%
  \BibitemOpen
  \bibfield  {author} {\bibinfo {author} {\bibfnamefont {S.}~\bibnamefont {Afolalu}}, \bibinfo {author} {\bibfnamefont {S.~B.}\ \bibnamefont {Soetan}}, \bibinfo {author} {\bibfnamefont {S.~O.}\ \bibnamefont {Ongbali}}, \bibinfo {author} {\bibfnamefont {A.}~\bibnamefont {Abioye}}, \ and\ \bibinfo {author} {\bibfnamefont {A.}~\bibnamefont {Oni}},\ }in\ \href@noop {} {\emph {\bibinfo {booktitle} {IOP Conference Series: Materials Science and Engineering}}},\ Vol.\ \bibinfo {volume} {640}\ (\bibinfo {organization} {IOP Publishing},\ \bibinfo {year} {2019})\ p.\ \bibinfo {pages} {012065}\BibitemShut {NoStop}%
\bibitem [{\citenamefont {Baig}\ \emph {et~al.}(2021)\citenamefont {Baig}, \citenamefont {Kammakakam},\ and\ \citenamefont {Falath}}]{baig2021nanomaterials}%
  \BibitemOpen
  \bibfield  {author} {\bibinfo {author} {\bibfnamefont {N.}~\bibnamefont {Baig}}, \bibinfo {author} {\bibfnamefont {I.}~\bibnamefont {Kammakakam}}, \ and\ \bibinfo {author} {\bibfnamefont {W.}~\bibnamefont {Falath}},\ }\href@noop {} {\bibfield  {journal} {\bibinfo  {journal} {Materials advances}\ }\textbf {\bibinfo {volume} {2}},\ \bibinfo {pages} {1821} (\bibinfo {year} {2021})}\BibitemShut {NoStop}%
\bibitem [{\citenamefont {Jiang}\ \emph {et~al.}(2020)\citenamefont {Jiang}, \citenamefont {Li}, \citenamefont {Shen}, \citenamefont {Shi}, \citenamefont {Lv}, \citenamefont {Zhang}, \citenamefont {Dong}, \citenamefont {Qi},\ and\ \citenamefont {Liu}}]{jiang2020mechanistic}%
  \BibitemOpen
  \bibfield  {author} {\bibinfo {author} {\bibfnamefont {G.}~\bibnamefont {Jiang}}, \bibinfo {author} {\bibfnamefont {X.}~\bibnamefont {Li}}, \bibinfo {author} {\bibfnamefont {Y.}~\bibnamefont {Shen}}, \bibinfo {author} {\bibfnamefont {X.}~\bibnamefont {Shi}}, \bibinfo {author} {\bibfnamefont {X.}~\bibnamefont {Lv}}, \bibinfo {author} {\bibfnamefont {X.}~\bibnamefont {Zhang}}, \bibinfo {author} {\bibfnamefont {F.}~\bibnamefont {Dong}}, \bibinfo {author} {\bibfnamefont {G.}~\bibnamefont {Qi}}, \ and\ \bibinfo {author} {\bibfnamefont {R.}~\bibnamefont {Liu}},\ }\href@noop {} {\bibfield  {journal} {\bibinfo  {journal} {Journal of Catalysis}\ }\textbf {\bibinfo {volume} {391}},\ \bibinfo {pages} {414} (\bibinfo {year} {2020})}\BibitemShut {NoStop}%
\bibitem [{\citenamefont {Guo}\ \emph {et~al.}(2023)\citenamefont {Guo}, \citenamefont {Gerstein}, \citenamefont {Jha}, \citenamefont {Arsano}, \citenamefont {Haider}, \citenamefont {Khan},\ and\ \citenamefont {Tsige}}]{facet_specific_adsorption}%
  \BibitemOpen
  \bibfield  {author} {\bibinfo {author} {\bibfnamefont {H.}~\bibnamefont {Guo}}, \bibinfo {author} {\bibfnamefont {E.~A.}\ \bibnamefont {Gerstein}}, \bibinfo {author} {\bibfnamefont {K.~C.}\ \bibnamefont {Jha}}, \bibinfo {author} {\bibfnamefont {I.}~\bibnamefont {Arsano}}, \bibinfo {author} {\bibfnamefont {M.~A.}\ \bibnamefont {Haider}}, \bibinfo {author} {\bibfnamefont {T.~S.}\ \bibnamefont {Khan}}, \ and\ \bibinfo {author} {\bibfnamefont {M.}~\bibnamefont {Tsige}},\ }\href@noop {} {\bibfield  {journal} {\bibinfo  {journal} {Frontiers in Catalysis}\ }\textbf {\bibinfo {volume} {3}},\ \bibinfo {pages} {1116867} (\bibinfo {year} {2023})}\BibitemShut {NoStop}%
\bibitem [{\citenamefont {Demiroglu}\ \emph {et~al.}(2016)\citenamefont {Demiroglu}, \citenamefont {Li}, \citenamefont {Piccolo},\ and\ \citenamefont {Johnston}}]{AuRh_clusters}%
  \BibitemOpen
  \bibfield  {author} {\bibinfo {author} {\bibfnamefont {I.}~\bibnamefont {Demiroglu}}, \bibinfo {author} {\bibfnamefont {Z.}~\bibnamefont {Li}}, \bibinfo {author} {\bibfnamefont {L.}~\bibnamefont {Piccolo}}, \ and\ \bibinfo {author} {\bibfnamefont {R.~L.}\ \bibnamefont {Johnston}},\ }\href@noop {} {\bibfield  {journal} {\bibinfo  {journal} {Catalysis Science \& Technology}\ }\textbf {\bibinfo {volume} {6}},\ \bibinfo {pages} {6916} (\bibinfo {year} {2016})}\BibitemShut {NoStop}%
\bibitem [{\citenamefont {Cao}\ \emph {et~al.}(2002)\citenamefont {Cao}, \citenamefont {Wang}, \citenamefont {Zhu}, \citenamefont {Wu},\ and\ \citenamefont {Zhang}}]{cao2002static}%
  \BibitemOpen
  \bibfield  {author} {\bibinfo {author} {\bibfnamefont {Z.}~\bibnamefont {Cao}}, \bibinfo {author} {\bibfnamefont {Y.}~\bibnamefont {Wang}}, \bibinfo {author} {\bibfnamefont {J.}~\bibnamefont {Zhu}}, \bibinfo {author} {\bibfnamefont {W.}~\bibnamefont {Wu}}, \ and\ \bibinfo {author} {\bibfnamefont {Q.}~\bibnamefont {Zhang}},\ }\href@noop {} {\bibfield  {journal} {\bibinfo  {journal} {The Journal of Physical Chemistry B}\ }\textbf {\bibinfo {volume} {106}},\ \bibinfo {pages} {9649} (\bibinfo {year} {2002})}\BibitemShut {NoStop}%
\bibitem [{\citenamefont {Zhang}\ and\ \citenamefont {Li}(2021)}]{Cu_cluster_sites}%
  \BibitemOpen
  \bibfield  {author} {\bibinfo {author} {\bibfnamefont {S.}~\bibnamefont {Zhang}}\ and\ \bibinfo {author} {\bibfnamefont {H.}~\bibnamefont {Li}},\ }\href@noop {} {\bibfield  {journal} {\bibinfo  {journal} {Vacuum}\ }\textbf {\bibinfo {volume} {184}},\ \bibinfo {pages} {109971} (\bibinfo {year} {2021})}\BibitemShut {NoStop}%
\bibitem [{\citenamefont {Ferrando}\ \emph {et~al.}(2008)\citenamefont {Ferrando}, \citenamefont {Jellinek},\ and\ \citenamefont {Johnston}}]{ferrando2008nanoalloys}%
  \BibitemOpen
  \bibfield  {author} {\bibinfo {author} {\bibfnamefont {R.}~\bibnamefont {Ferrando}}, \bibinfo {author} {\bibfnamefont {J.}~\bibnamefont {Jellinek}}, \ and\ \bibinfo {author} {\bibfnamefont {R.~L.}\ \bibnamefont {Johnston}},\ }\href@noop {} {\bibfield  {journal} {\bibinfo  {journal} {Chemical reviews}\ }\textbf {\bibinfo {volume} {108}},\ \bibinfo {pages} {845} (\bibinfo {year} {2008})}\BibitemShut {NoStop}%
\bibitem [{\citenamefont {Makarucha}\ \emph {et~al.}(2011)\citenamefont {Makarucha}, \citenamefont {Todorova},\ and\ \citenamefont {Yarovsky}}]{bio-nanomaterials}%
  \BibitemOpen
  \bibfield  {author} {\bibinfo {author} {\bibfnamefont {A.}~\bibnamefont {Makarucha}}, \bibinfo {author} {\bibfnamefont {N.}~\bibnamefont {Todorova}}, \ and\ \bibinfo {author} {\bibfnamefont {I.}~\bibnamefont {Yarovsky}},\ }\href@noop {} {\bibfield  {journal} {\bibinfo  {journal} {European Biophysics Journal}\ }\textbf {\bibinfo {volume} {40}},\ \bibinfo {pages} {103} (\bibinfo {year} {2011})}\BibitemShut {NoStop}%
\bibitem [{\citenamefont {Calvo}(2015)}]{calvo2015thermodynamics}%
  \BibitemOpen
  \bibfield  {author} {\bibinfo {author} {\bibfnamefont {F.}~\bibnamefont {Calvo}},\ }\href@noop {} {\bibfield  {journal} {\bibinfo  {journal} {Physical Chemistry Chemical Physics}\ }\textbf {\bibinfo {volume} {17}},\ \bibinfo {pages} {27922} (\bibinfo {year} {2015})}\BibitemShut {NoStop}%
\bibitem [{\citenamefont {Pidko}(2017)}]{pidko2017toward}%
  \BibitemOpen
  \bibfield  {author} {\bibinfo {author} {\bibfnamefont {E.~A.}\ \bibnamefont {Pidko}},\ }\href@noop {} {\bibfield  {journal} {\bibinfo  {journal} {ACS Catalysis}\ }\textbf {\bibinfo {volume} {7}},\ \bibinfo {pages} {4230} (\bibinfo {year} {2017})}\BibitemShut {NoStop}%
\bibitem [{\citenamefont {Douglas-Gallardo}\ \emph {et~al.}(2021)\citenamefont {Douglas-Gallardo}, \citenamefont {Box},\ and\ \citenamefont {Maurer}}]{douglas2021plasmonic}%
  \BibitemOpen
  \bibfield  {author} {\bibinfo {author} {\bibfnamefont {O.~A.}\ \bibnamefont {Douglas-Gallardo}}, \bibinfo {author} {\bibfnamefont {C.~L.}\ \bibnamefont {Box}}, \ and\ \bibinfo {author} {\bibfnamefont {R.~J.}\ \bibnamefont {Maurer}},\ }\href@noop {} {\bibfield  {journal} {\bibinfo  {journal} {Nanoscale}\ }\textbf {\bibinfo {volume} {13}},\ \bibinfo {pages} {11058} (\bibinfo {year} {2021})}\BibitemShut {NoStop}%
\bibitem [{\citenamefont {Catlow}(2020)}]{catlow2020computational}%
  \BibitemOpen
  \bibfield  {author} {\bibinfo {author} {\bibfnamefont {C.~R.~A.}\ \bibnamefont {Catlow}},\ }\href@noop {} {\enquote {\bibinfo {title} {Computational modelling as a tool in structural science},}\ } (\bibinfo {year} {2020})\BibitemShut {NoStop}%
\bibitem [{\citenamefont {Grajciar}\ \emph {et~al.}(2018)\citenamefont {Grajciar}, \citenamefont {Heard}, \citenamefont {Bondarenko}, \citenamefont {Polynski}, \citenamefont {Meeprasert}, \citenamefont {Pidko},\ and\ \citenamefont {Nachtigall}}]{grajciar2018towards}%
  \BibitemOpen
  \bibfield  {author} {\bibinfo {author} {\bibfnamefont {L.}~\bibnamefont {Grajciar}}, \bibinfo {author} {\bibfnamefont {C.~J.}\ \bibnamefont {Heard}}, \bibinfo {author} {\bibfnamefont {A.~A.}\ \bibnamefont {Bondarenko}}, \bibinfo {author} {\bibfnamefont {M.~V.}\ \bibnamefont {Polynski}}, \bibinfo {author} {\bibfnamefont {J.}~\bibnamefont {Meeprasert}}, \bibinfo {author} {\bibfnamefont {E.~A.}\ \bibnamefont {Pidko}}, \ and\ \bibinfo {author} {\bibfnamefont {P.}~\bibnamefont {Nachtigall}},\ }\href@noop {} {\bibfield  {journal} {\bibinfo  {journal} {Chemical Society Reviews}\ }\textbf {\bibinfo {volume} {47}},\ \bibinfo {pages} {8307} (\bibinfo {year} {2018})}\BibitemShut {NoStop}%
\bibitem [{\citenamefont {Wales}\ and\ \citenamefont {Doye}(1997{\natexlab{a}})}]{wales1997global}%
  \BibitemOpen
  \bibfield  {author} {\bibinfo {author} {\bibfnamefont {D.~J.}\ \bibnamefont {Wales}}\ and\ \bibinfo {author} {\bibfnamefont {J.~P.}\ \bibnamefont {Doye}},\ }\href@noop {} {\bibfield  {journal} {\bibinfo  {journal} {The Journal of Physical Chemistry A}\ }\textbf {\bibinfo {volume} {101}},\ \bibinfo {pages} {5111} (\bibinfo {year} {1997}{\natexlab{a}})}\BibitemShut {NoStop}%
\bibitem [{\citenamefont {Lazauskas}\ \emph {et~al.}(2017)\citenamefont {Lazauskas}, \citenamefont {Sokol},\ and\ \citenamefont {Woodley}}]{lazauskas2017efficient}%
  \BibitemOpen
  \bibfield  {author} {\bibinfo {author} {\bibfnamefont {T.}~\bibnamefont {Lazauskas}}, \bibinfo {author} {\bibfnamefont {A.~A.}\ \bibnamefont {Sokol}}, \ and\ \bibinfo {author} {\bibfnamefont {S.~M.}\ \bibnamefont {Woodley}},\ }\href@noop {} {\bibfield  {journal} {\bibinfo  {journal} {Nanoscale}\ }\textbf {\bibinfo {volume} {9}},\ \bibinfo {pages} {3850} (\bibinfo {year} {2017})}\BibitemShut {NoStop}%
\bibitem [{\citenamefont {Woodley}\ \emph {et~al.}(2020)\citenamefont {Woodley}, \citenamefont {Day},\ and\ \citenamefont {Catlow}}]{woodley2020structure}%
  \BibitemOpen
  \bibfield  {author} {\bibinfo {author} {\bibfnamefont {S.~M.}\ \bibnamefont {Woodley}}, \bibinfo {author} {\bibfnamefont {G.~M.}\ \bibnamefont {Day}}, \ and\ \bibinfo {author} {\bibfnamefont {R.}~\bibnamefont {Catlow}},\ }\href@noop {} {\bibfield  {journal} {\bibinfo  {journal} {Philosophical Transactions of the Royal Society A}\ }\textbf {\bibinfo {volume} {378}},\ \bibinfo {pages} {20190600} (\bibinfo {year} {2020})}\BibitemShut {NoStop}%
\bibitem [{\citenamefont {Nabi}\ \emph {et~al.}(2022)\citenamefont {Nabi}, \citenamefont {Hussain}, \citenamefont {Di~Tommaso} \emph {et~al.}}]{nabi2022ab}%
  \BibitemOpen
  \bibfield  {author} {\bibinfo {author} {\bibfnamefont {A.~G.}\ \bibnamefont {Nabi}}, \bibinfo {author} {\bibfnamefont {A.}~\bibnamefont {Hussain}}, \bibinfo {author} {\bibfnamefont {D.}~\bibnamefont {Di~Tommaso}},  \emph {et~al.},\ }\href@noop {} {\bibfield  {journal} {\bibinfo  {journal} {Molecular Catalysis}\ }\textbf {\bibinfo {volume} {527}},\ \bibinfo {pages} {112406} (\bibinfo {year} {2022})}\BibitemShut {NoStop}%
\bibitem [{\citenamefont {Baletto}\ \emph {et~al.}(2002)\citenamefont {Baletto}, \citenamefont {Ferrando}, \citenamefont {Fortunelli}, \citenamefont {Montalenti},\ and\ \citenamefont {Mottet}}]{baletto2002crossover}%
  \BibitemOpen
  \bibfield  {author} {\bibinfo {author} {\bibfnamefont {F.}~\bibnamefont {Baletto}}, \bibinfo {author} {\bibfnamefont {R.}~\bibnamefont {Ferrando}}, \bibinfo {author} {\bibfnamefont {A.}~\bibnamefont {Fortunelli}}, \bibinfo {author} {\bibfnamefont {F.}~\bibnamefont {Montalenti}}, \ and\ \bibinfo {author} {\bibfnamefont {C.}~\bibnamefont {Mottet}},\ }\href@noop {} {\bibfield  {journal} {\bibinfo  {journal} {J.~Chem.~Phys.}\ }\textbf {\bibinfo {volume} {116}},\ \bibinfo {pages} {3856} (\bibinfo {year} {2002})}\BibitemShut {NoStop}%
\bibitem [{\citenamefont {Rossi}\ \emph {et~al.}(2018)\citenamefont {Rossi}, \citenamefont {P\'artay}, \citenamefont {Cs\'anyi},\ and\ \citenamefont {Baletto}}]{CuPt_ns}%
  \BibitemOpen
  \bibfield  {author} {\bibinfo {author} {\bibfnamefont {K.}~\bibnamefont {Rossi}}, \bibinfo {author} {\bibfnamefont {L.}~\bibnamefont {P\'artay}}, \bibinfo {author} {\bibfnamefont {G.}~\bibnamefont {Cs\'anyi}}, \ and\ \bibinfo {author} {\bibfnamefont {F.}~\bibnamefont {Baletto}},\ }\href@noop {} {\bibfield  {journal} {\bibinfo  {journal} {Scientific Reports}\ }\textbf {\bibinfo {volume} {8}},\ \bibinfo {pages} {9150} (\bibinfo {year} {2018})}\BibitemShut {NoStop}%
\bibitem [{\citenamefont {Casey-Stevens}\ \emph {et~al.}(2021)\citenamefont {Casey-Stevens}, \citenamefont {Yang}, \citenamefont {Weal}, \citenamefont {McIntyre}, \citenamefont {Nally},\ and\ \citenamefont {Garden}}]{casey2021theoretical}%
  \BibitemOpen
  \bibfield  {author} {\bibinfo {author} {\bibfnamefont {C.~A.}\ \bibnamefont {Casey-Stevens}}, \bibinfo {author} {\bibfnamefont {M.}~\bibnamefont {Yang}}, \bibinfo {author} {\bibfnamefont {G.~R.}\ \bibnamefont {Weal}}, \bibinfo {author} {\bibfnamefont {S.~M.}\ \bibnamefont {McIntyre}}, \bibinfo {author} {\bibfnamefont {B.~K.}\ \bibnamefont {Nally}}, \ and\ \bibinfo {author} {\bibfnamefont {A.~L.}\ \bibnamefont {Garden}},\ }\href@noop {} {\bibfield  {journal} {\bibinfo  {journal} {Physical Chemistry Chemical Physics}\ }\textbf {\bibinfo {volume} {23}},\ \bibinfo {pages} {15950} (\bibinfo {year} {2021})}\BibitemShut {NoStop}%
\bibitem [{\citenamefont {Borg}\ \emph {et~al.}(2005)\citenamefont {Borg}, \citenamefont {Stampfl}, \citenamefont {Mikkelsen}, \citenamefont {Gustafson}, \citenamefont {Lundgren}, \citenamefont {Scheffler},\ and\ \citenamefont {Andersen}}]{borg2005density}%
  \BibitemOpen
  \bibfield  {author} {\bibinfo {author} {\bibfnamefont {M.}~\bibnamefont {Borg}}, \bibinfo {author} {\bibfnamefont {C.}~\bibnamefont {Stampfl}}, \bibinfo {author} {\bibfnamefont {A.}~\bibnamefont {Mikkelsen}}, \bibinfo {author} {\bibfnamefont {J.}~\bibnamefont {Gustafson}}, \bibinfo {author} {\bibfnamefont {E.}~\bibnamefont {Lundgren}}, \bibinfo {author} {\bibfnamefont {M.}~\bibnamefont {Scheffler}}, \ and\ \bibinfo {author} {\bibfnamefont {J.~N.}\ \bibnamefont {Andersen}},\ }\href@noop {} {\bibfield  {journal} {\bibinfo  {journal} {ChemPhysChem}\ }\textbf {\bibinfo {volume} {6}},\ \bibinfo {pages} {1923} (\bibinfo {year} {2005})}\BibitemShut {NoStop}%
\bibitem [{\citenamefont {Reuter}\ and\ \citenamefont {Scheffler}(2001)}]{reuter2001composition}%
  \BibitemOpen
  \bibfield  {author} {\bibinfo {author} {\bibfnamefont {K.}~\bibnamefont {Reuter}}\ and\ \bibinfo {author} {\bibfnamefont {M.}~\bibnamefont {Scheffler}},\ }\href@noop {} {\bibfield  {journal} {\bibinfo  {journal} {Physical Review B}\ }\textbf {\bibinfo {volume} {65}},\ \bibinfo {pages} {035406} (\bibinfo {year} {2001})}\BibitemShut {NoStop}%
\bibitem [{\citenamefont {Stampfl}\ \emph {et~al.}(2002)\citenamefont {Stampfl}, \citenamefont {Ganduglia-Pirovano}, \citenamefont {Reuter},\ and\ \citenamefont {Scheffler}}]{stampfl2002catalysis}%
  \BibitemOpen
  \bibfield  {author} {\bibinfo {author} {\bibfnamefont {C.}~\bibnamefont {Stampfl}}, \bibinfo {author} {\bibfnamefont {M.~V.}\ \bibnamefont {Ganduglia-Pirovano}}, \bibinfo {author} {\bibfnamefont {K.}~\bibnamefont {Reuter}}, \ and\ \bibinfo {author} {\bibfnamefont {M.}~\bibnamefont {Scheffler}},\ }\href@noop {} {\bibfield  {journal} {\bibinfo  {journal} {Surface Science}\ }\textbf {\bibinfo {volume} {500}},\ \bibinfo {pages} {368} (\bibinfo {year} {2002})}\BibitemShut {NoStop}%
\bibitem [{\citenamefont {Lee}\ and\ \citenamefont {Soon}(2024)}]{lee2024rise}%
  \BibitemOpen
  \bibfield  {author} {\bibinfo {author} {\bibfnamefont {T.}~\bibnamefont {Lee}}\ and\ \bibinfo {author} {\bibfnamefont {A.}~\bibnamefont {Soon}},\ }\href@noop {} {\bibfield  {journal} {\bibinfo  {journal} {Nature Catalysis}\ }\textbf {\bibinfo {volume} {7}},\ \bibinfo {pages} {4} (\bibinfo {year} {2024})}\BibitemShut {NoStop}%
\bibitem [{\citenamefont {Wexler}\ \emph {et~al.}(2019)\citenamefont {Wexler}, \citenamefont {Qiu},\ and\ \citenamefont {Rappe}}]{wexler2019automatic}%
  \BibitemOpen
  \bibfield  {author} {\bibinfo {author} {\bibfnamefont {R.~B.}\ \bibnamefont {Wexler}}, \bibinfo {author} {\bibfnamefont {T.}~\bibnamefont {Qiu}}, \ and\ \bibinfo {author} {\bibfnamefont {A.~M.}\ \bibnamefont {Rappe}},\ }\href@noop {} {\bibfield  {journal} {\bibinfo  {journal} {The Journal of Physical Chemistry C}\ }\textbf {\bibinfo {volume} {123}},\ \bibinfo {pages} {2321} (\bibinfo {year} {2019})}\BibitemShut {NoStop}%
\bibitem [{\citenamefont {Dean}\ \emph {et~al.}(2019)\citenamefont {Dean}, \citenamefont {Taylor},\ and\ \citenamefont {Mpourmpakis}}]{MLadsorption_model}%
  \BibitemOpen
  \bibfield  {author} {\bibinfo {author} {\bibfnamefont {J.}~\bibnamefont {Dean}}, \bibinfo {author} {\bibfnamefont {M.~G.}\ \bibnamefont {Taylor}}, \ and\ \bibinfo {author} {\bibfnamefont {G.}~\bibnamefont {Mpourmpakis}},\ }\href@noop {} {\bibfield  {journal} {\bibinfo  {journal} {Science advances}\ }\textbf {\bibinfo {volume} {5}},\ \bibinfo {pages} {eaax5101} (\bibinfo {year} {2019})}\BibitemShut {NoStop}%
\bibitem [{\citenamefont {Zhou}\ \emph {et~al.}(2019)\citenamefont {Zhou}, \citenamefont {Scheffler},\ and\ \citenamefont {Ghiringhelli}}]{zhou2019determining}%
  \BibitemOpen
  \bibfield  {author} {\bibinfo {author} {\bibfnamefont {Y.}~\bibnamefont {Zhou}}, \bibinfo {author} {\bibfnamefont {M.}~\bibnamefont {Scheffler}}, \ and\ \bibinfo {author} {\bibfnamefont {L.~M.}\ \bibnamefont {Ghiringhelli}},\ }\href@noop {} {\bibfield  {journal} {\bibinfo  {journal} {Physical Review B}\ }\textbf {\bibinfo {volume} {100}},\ \bibinfo {pages} {174106} (\bibinfo {year} {2019})}\BibitemShut {NoStop}%
\bibitem [{\citenamefont {P\'artay}\ \emph {et~al.}(2010)\citenamefont {P\'artay}, \citenamefont {Bart\'ok},\ and\ \citenamefont {Cs\'anyi}}]{1st_NS_paper}%
  \BibitemOpen
  \bibfield  {author} {\bibinfo {author} {\bibfnamefont {L.~B.}\ \bibnamefont {P\'artay}}, \bibinfo {author} {\bibfnamefont {A.~P.}\ \bibnamefont {Bart\'ok}}, \ and\ \bibinfo {author} {\bibfnamefont {G.}~\bibnamefont {Cs\'anyi}},\ }\href@noop {} {\bibfield  {journal} {\bibinfo  {journal} {J. Phys. Chem. B}\ }\textbf {\bibinfo {volume} {114}},\ \bibinfo {pages} {10502} (\bibinfo {year} {2010})}\BibitemShut {NoStop}%
\bibitem [{\citenamefont {Yang}\ \emph {et~al.}(2024)\citenamefont {Yang}, \citenamefont {Pártay},\ and\ \citenamefont {Wexler}}]{yang2024surface}%
  \BibitemOpen
  \bibfield  {author} {\bibinfo {author} {\bibfnamefont {M.}~\bibnamefont {Yang}}, \bibinfo {author} {\bibfnamefont {L.~B.}\ \bibnamefont {Pártay}}, \ and\ \bibinfo {author} {\bibfnamefont {R.~B.}\ \bibnamefont {Wexler}},\ }\href@noop {} {\bibfield  {journal} {\bibinfo  {journal} {Phys, Chem. Chem. Phys.}\ }\textbf {\bibinfo {volume} {26}},\ \bibinfo {pages} {13862} (\bibinfo {year} {2024})}\BibitemShut {NoStop}%
\bibitem [{\citenamefont {Wales}\ and\ \citenamefont {Doye}(1997{\natexlab{b}})}]{bib:wales_basin_LJ}%
  \BibitemOpen
  \bibfield  {author} {\bibinfo {author} {\bibfnamefont {D.~J.}\ \bibnamefont {Wales}}\ and\ \bibinfo {author} {\bibfnamefont {J.~P.~K.}\ \bibnamefont {Doye}},\ }\href@noop {} {\bibfield  {journal} {\bibinfo  {journal} {J. Phys. Chem. A}\ }\textbf {\bibinfo {volume} {101}},\ \bibinfo {pages} {5111} (\bibinfo {year} {1997}{\natexlab{b}})}\BibitemShut {NoStop}%
\bibitem [{\citenamefont {Skilling}(2004)}]{bib:skilling}%
  \BibitemOpen
  \bibfield  {author} {\bibinfo {author} {\bibfnamefont {J.}~\bibnamefont {Skilling}},\ }in\ \href@noop {} {\emph {\bibinfo {booktitle} {AIP Conference Proceedings}}},\ Vol.\ \bibinfo {volume} {735}\ (\bibinfo {year} {2004})\ p.\ \bibinfo {pages} {395}\BibitemShut {NoStop}%
\bibitem [{\citenamefont {Skilling}(2006)}]{bib:skilling2}%
  \BibitemOpen
  \bibfield  {author} {\bibinfo {author} {\bibfnamefont {J.}~\bibnamefont {Skilling}},\ }\href@noop {} {\bibfield  {journal} {\bibinfo  {journal} {J. of Bayesian Analysis}\ }\textbf {\bibinfo {volume} {1}},\ \bibinfo {pages} {833} (\bibinfo {year} {2006})}\BibitemShut {NoStop}%
\bibitem [{\citenamefont {Ashton}\ \emph {et~al.}(2022)\citenamefont {Ashton}, \citenamefont {Bernstein}, \citenamefont {Buchner}, \citenamefont {Chen}, \citenamefont {Csányi}, \citenamefont {Feroz}, \citenamefont {Fowlie}, \citenamefont {Griffiths}, \citenamefont {Habeck}, \citenamefont {Handley}, \citenamefont {Higson}, \citenamefont {Hobson}, \citenamefont {Lasenby}, \citenamefont {Parkinson}, \citenamefont {Pártay}, \citenamefont {Pitkin}, \citenamefont {Schneider}, \citenamefont {South}, \citenamefont {Speagle}, \citenamefont {Veitch}, \citenamefont {Wacker}, \citenamefont {Wales},\ and\ \citenamefont {Yallup}}]{NS_all_review}%
  \BibitemOpen
  \bibfield  {author} {\bibinfo {author} {\bibfnamefont {G.}~\bibnamefont {Ashton}}, \bibinfo {author} {\bibfnamefont {N.}~\bibnamefont {Bernstein}}, \bibinfo {author} {\bibfnamefont {J.}~\bibnamefont {Buchner}}, \bibinfo {author} {\bibfnamefont {X.}~\bibnamefont {Chen}}, \bibinfo {author} {\bibfnamefont {G.}~\bibnamefont {Csányi}}, \bibinfo {author} {\bibfnamefont {F.}~\bibnamefont {Feroz}}, \bibinfo {author} {\bibfnamefont {A.}~\bibnamefont {Fowlie}}, \bibinfo {author} {\bibfnamefont {M.}~\bibnamefont {Griffiths}}, \bibinfo {author} {\bibfnamefont {M.}~\bibnamefont {Habeck}}, \bibinfo {author} {\bibfnamefont {W.}~\bibnamefont {Handley}}, \bibinfo {author} {\bibfnamefont {E.}~\bibnamefont {Higson}}, \bibinfo {author} {\bibfnamefont {M.}~\bibnamefont {Hobson}}, \bibinfo {author} {\bibfnamefont {A.}~\bibnamefont {Lasenby}}, \bibinfo {author} {\bibfnamefont {D.}~\bibnamefont {Parkinson}}, \bibinfo {author} {\bibfnamefont {L.~B.}\ \bibnamefont {Pártay}}, \bibinfo {author} {\bibfnamefont {M.}~\bibnamefont
  {Pitkin}}, \bibinfo {author} {\bibfnamefont {D.}~\bibnamefont {Schneider}}, \bibinfo {author} {\bibfnamefont {L.}~\bibnamefont {South}}, \bibinfo {author} {\bibfnamefont {J.~S.}\ \bibnamefont {Speagle}}, \bibinfo {author} {\bibfnamefont {J.}~\bibnamefont {Veitch}}, \bibinfo {author} {\bibfnamefont {P.}~\bibnamefont {Wacker}}, \bibinfo {author} {\bibfnamefont {D.~J.}\ \bibnamefont {Wales}}, \ and\ \bibinfo {author} {\bibfnamefont {D.}~\bibnamefont {Yallup}},\ }\href@noop {} {\bibfield  {journal} {\bibinfo  {journal} {Nat. Rev. Methods Primer}\ }\textbf {\bibinfo {volume} {2}},\ \bibinfo {pages} {39} (\bibinfo {year} {2022})}\BibitemShut {NoStop}%
\bibitem [{\citenamefont {P\'artay}\ \emph {et~al.}(2021)\citenamefont {P\'artay}, \citenamefont {Cs\'anyi},\ and\ \citenamefont {Bernstein}}]{NS_mat_review}%
  \BibitemOpen
  \bibfield  {author} {\bibinfo {author} {\bibfnamefont {L.~B.}\ \bibnamefont {P\'artay}}, \bibinfo {author} {\bibfnamefont {G.}~\bibnamefont {Cs\'anyi}}, \ and\ \bibinfo {author} {\bibfnamefont {N.}~\bibnamefont {Bernstein}},\ }\href@noop {} {\bibfield  {journal} {\bibinfo  {journal} {Eur. Phys. J. B}\ }\textbf {\bibinfo {volume} {94}},\ \bibinfo {pages} {159} (\bibinfo {year} {2021})}\BibitemShut {NoStop}%
\bibitem [{\citenamefont {Martiniani}\ \emph {et~al.}(2014)\citenamefont {Martiniani}, \citenamefont {Stevenson}, \citenamefont {Wales},\ and\ \citenamefont {Frenkel}}]{Frenkel_NS}%
  \BibitemOpen
  \bibfield  {author} {\bibinfo {author} {\bibfnamefont {S.}~\bibnamefont {Martiniani}}, \bibinfo {author} {\bibfnamefont {J.~D.}\ \bibnamefont {Stevenson}}, \bibinfo {author} {\bibfnamefont {D.~J.}\ \bibnamefont {Wales}}, \ and\ \bibinfo {author} {\bibfnamefont {D.}~\bibnamefont {Frenkel}},\ }\href@noop {} {\bibfield  {journal} {\bibinfo  {journal} {Phys. Rev. X}\ }\textbf {\bibinfo {volume} {4}},\ \bibinfo {pages} {031034} (\bibinfo {year} {2014})}\BibitemShut {NoStop}%
\bibitem [{\citenamefont {Baldock}\ \emph {et~al.}(2016)\citenamefont {Baldock}, \citenamefont {P\'artay}, \citenamefont {Bart\'ok}, \citenamefont {Payne},\ and\ \citenamefont {Cs\'anyi}}]{pt_phase_dias_ns}%
  \BibitemOpen
  \bibfield  {author} {\bibinfo {author} {\bibfnamefont {R.~J.~N.}\ \bibnamefont {Baldock}}, \bibinfo {author} {\bibfnamefont {L.~B.}\ \bibnamefont {P\'artay}}, \bibinfo {author} {\bibfnamefont {A.~P.}\ \bibnamefont {Bart\'ok}}, \bibinfo {author} {\bibfnamefont {M.~C.}\ \bibnamefont {Payne}}, \ and\ \bibinfo {author} {\bibfnamefont {G.}~\bibnamefont {Cs\'anyi}},\ }\href {\doibase 10.1103/PhysRevB.93.174108} {\bibfield  {journal} {\bibinfo  {journal} {Phys. Rev. B}\ }\textbf {\bibinfo {volume} {93}},\ \bibinfo {pages} {174108} (\bibinfo {year} {2016})}\BibitemShut {NoStop}%
\bibitem [{\citenamefont {Baldock}\ \emph {et~al.}(2017)\citenamefont {Baldock}, \citenamefont {Bernstein}, \citenamefont {Salerno}, \citenamefont {P\'artay},\ and\ \citenamefont {Cs\'anyi}}]{ConPresNS}%
  \BibitemOpen
  \bibfield  {author} {\bibinfo {author} {\bibfnamefont {R.~J.~N.}\ \bibnamefont {Baldock}}, \bibinfo {author} {\bibfnamefont {N.}~\bibnamefont {Bernstein}}, \bibinfo {author} {\bibfnamefont {K.~M.}\ \bibnamefont {Salerno}}, \bibinfo {author} {\bibfnamefont {L.~B.}\ \bibnamefont {P\'artay}}, \ and\ \bibinfo {author} {\bibfnamefont {G.}~\bibnamefont {Cs\'anyi}},\ }\href@noop {} {\bibfield  {journal} {\bibinfo  {journal} {Phys. Rev. E}\ }\textbf {\bibinfo {volume} {96}},\ \bibinfo {pages} {43311} (\bibinfo {year} {2017})}\BibitemShut {NoStop}%
\bibitem [{\citenamefont {Bichoutskaia}\ and\ \citenamefont {Pyper}(2008)}]{bichoutskaia2008}%
  \BibitemOpen
  \bibfield  {author} {\bibinfo {author} {\bibfnamefont {E.}~\bibnamefont {Bichoutskaia}}\ and\ \bibinfo {author} {\bibfnamefont {N.~C.}\ \bibnamefont {Pyper}},\ }\href@noop {} {\bibfield  {journal} {\bibinfo  {journal} {J.~Chem.~Phys.}\ }\textbf {\bibinfo {volume} {128}} (\bibinfo {year} {2008})}\BibitemShut {NoStop}%
\bibitem [{\citenamefont {Miasojedow}\ \emph {et~al.}(2013)\citenamefont {Miasojedow}, \citenamefont {Moulines},\ and\ \citenamefont {Vihola}}]{miasojedow2013adaptive}%
  \BibitemOpen
  \bibfield  {author} {\bibinfo {author} {\bibfnamefont {B.}~\bibnamefont {Miasojedow}}, \bibinfo {author} {\bibfnamefont {E.}~\bibnamefont {Moulines}}, \ and\ \bibinfo {author} {\bibfnamefont {M.}~\bibnamefont {Vihola}},\ }\href@noop {} {\bibfield  {journal} {\bibinfo  {journal} {Journal of Computational and Graphical Statistics}\ }\textbf {\bibinfo {volume} {22}},\ \bibinfo {pages} {649} (\bibinfo {year} {2013})}\BibitemShut {NoStop}%
\bibitem [{\citenamefont {Rozada}\ \emph {et~al.}(2019)\citenamefont {Rozada}, \citenamefont {Aramon}, \citenamefont {Machta},\ and\ \citenamefont {Katzgraber}}]{rozada2019effects}%
  \BibitemOpen
  \bibfield  {author} {\bibinfo {author} {\bibfnamefont {I.}~\bibnamefont {Rozada}}, \bibinfo {author} {\bibfnamefont {M.}~\bibnamefont {Aramon}}, \bibinfo {author} {\bibfnamefont {J.}~\bibnamefont {Machta}}, \ and\ \bibinfo {author} {\bibfnamefont {H.~G.}\ \bibnamefont {Katzgraber}},\ }\href@noop {} {\bibfield  {journal} {\bibinfo  {journal} {Physical Review E}\ }\textbf {\bibinfo {volume} {100}},\ \bibinfo {pages} {043311} (\bibinfo {year} {2019})}\BibitemShut {NoStop}%
\bibitem [{\citenamefont {Terzyk}\ \emph {et~al.}(2007)\citenamefont {Terzyk}, \citenamefont {Furmaniak}, \citenamefont {Gauden}, \citenamefont {Harris}, \citenamefont {W{\l}och},\ and\ \citenamefont {Kowalczyk}}]{terzyk2007hyper}%
  \BibitemOpen
  \bibfield  {author} {\bibinfo {author} {\bibfnamefont {A.~P.}\ \bibnamefont {Terzyk}}, \bibinfo {author} {\bibfnamefont {S.}~\bibnamefont {Furmaniak}}, \bibinfo {author} {\bibfnamefont {P.~A.}\ \bibnamefont {Gauden}}, \bibinfo {author} {\bibfnamefont {P.~J.}\ \bibnamefont {Harris}}, \bibinfo {author} {\bibfnamefont {J.}~\bibnamefont {W{\l}och}}, \ and\ \bibinfo {author} {\bibfnamefont {P.}~\bibnamefont {Kowalczyk}},\ }\href@noop {} {\bibfield  {journal} {\bibinfo  {journal} {Journal of Physics: Condensed Matter}\ }\textbf {\bibinfo {volume} {19}},\ \bibinfo {pages} {406208} (\bibinfo {year} {2007})}\BibitemShut {NoStop}%
\bibitem [{\citenamefont {Liewehr}\ and\ \citenamefont {Bachmann}(2016)}]{liewehr2016homopolymer}%
  \BibitemOpen
  \bibfield  {author} {\bibinfo {author} {\bibfnamefont {B.}~\bibnamefont {Liewehr}}\ and\ \bibinfo {author} {\bibfnamefont {M.}~\bibnamefont {Bachmann}},\ }in\ \href@noop {} {\emph {\bibinfo {booktitle} {Journal of Physics: Conference Series}}},\ Vol.\ \bibinfo {volume} {686}\ (\bibinfo {organization} {IOP Publishing},\ \bibinfo {year} {2016})\ p.\ \bibinfo {pages} {012002}\BibitemShut {NoStop}%
\bibitem [{\citenamefont {Xie}\ \emph {et~al.}(2018)\citenamefont {Xie}, \citenamefont {Li},\ and\ \citenamefont {Zhou}}]{xie2018hamiltonian}%
  \BibitemOpen
  \bibfield  {author} {\bibinfo {author} {\bibfnamefont {Y.}~\bibnamefont {Xie}}, \bibinfo {author} {\bibfnamefont {Z.}~\bibnamefont {Li}}, \ and\ \bibinfo {author} {\bibfnamefont {J.}~\bibnamefont {Zhou}},\ }\href@noop {} {\bibfield  {journal} {\bibinfo  {journal} {Physical Chemistry Chemical Physics}\ }\textbf {\bibinfo {volume} {20}},\ \bibinfo {pages} {14587} (\bibinfo {year} {2018})}\BibitemShut {NoStop}%
\bibitem [{\citenamefont {Mehendale}\ \emph {et~al.}(2025)\citenamefont {Mehendale}, \citenamefont {Vlachos},\ and\ \citenamefont {Caratzoulas}}]{mehendale2025effect}%
  \BibitemOpen
  \bibfield  {author} {\bibinfo {author} {\bibfnamefont {R.~M.}\ \bibnamefont {Mehendale}}, \bibinfo {author} {\bibfnamefont {D.~G.}\ \bibnamefont {Vlachos}}, \ and\ \bibinfo {author} {\bibfnamefont {S.}~\bibnamefont {Caratzoulas}},\ }\href@noop {} {\bibfield  {journal} {\bibinfo  {journal} {The Journal of Physical Chemistry C}\ } (\bibinfo {year} {2025})}\BibitemShut {NoStop}%
\bibitem [{\citenamefont {Ciobanu}\ and\ \citenamefont {Predescu}(2004)}]{ciobanu2004reconstruction}%
  \BibitemOpen
  \bibfield  {author} {\bibinfo {author} {\bibfnamefont {C.~V.}\ \bibnamefont {Ciobanu}}\ and\ \bibinfo {author} {\bibfnamefont {C.}~\bibnamefont {Predescu}},\ }\href@noop {} {\bibfield  {journal} {\bibinfo  {journal} {Physical Review B—Condensed Matter and Materials Physics}\ }\textbf {\bibinfo {volume} {70}},\ \bibinfo {pages} {085321} (\bibinfo {year} {2004})}\BibitemShut {NoStop}%
\bibitem [{\citenamefont {Thompson}\ \emph {et~al.}(2022)\citenamefont {Thompson}, \citenamefont {Aktulga}, \citenamefont {Berger}, \citenamefont {Bolintineanu}, \citenamefont {Brown}, \citenamefont {Crozier}, \citenamefont {In't~Veld}, \citenamefont {Kohlmeyer}, \citenamefont {Moore}, \citenamefont {Nguyen} \emph {et~al.}}]{t2022lammps}%
  \BibitemOpen
  \bibfield  {author} {\bibinfo {author} {\bibfnamefont {A.~P.}\ \bibnamefont {Thompson}}, \bibinfo {author} {\bibfnamefont {H.~M.}\ \bibnamefont {Aktulga}}, \bibinfo {author} {\bibfnamefont {R.}~\bibnamefont {Berger}}, \bibinfo {author} {\bibfnamefont {D.~S.}\ \bibnamefont {Bolintineanu}}, \bibinfo {author} {\bibfnamefont {W.~M.}\ \bibnamefont {Brown}}, \bibinfo {author} {\bibfnamefont {P.~S.}\ \bibnamefont {Crozier}}, \bibinfo {author} {\bibfnamefont {P.~J.}\ \bibnamefont {In't~Veld}}, \bibinfo {author} {\bibfnamefont {A.}~\bibnamefont {Kohlmeyer}}, \bibinfo {author} {\bibfnamefont {S.~G.}\ \bibnamefont {Moore}}, \bibinfo {author} {\bibfnamefont {T.~D.}\ \bibnamefont {Nguyen}},  \emph {et~al.},\ }\href@noop {} {\bibfield  {journal} {\bibinfo  {journal} {Computer physics communications}\ }\textbf {\bibinfo {volume} {271}},\ \bibinfo {pages} {108171} (\bibinfo {year} {2022})}\BibitemShut {NoStop}%
\bibitem [{\citenamefont {Chatbipho}\ \emph {et~al.}(2025)\citenamefont {Chatbipho}, \citenamefont {Yang}, \citenamefont {Wexler},\ and\ \citenamefont {Partay}}]{chatbipho_adsorbate_2025}%
  \BibitemOpen
  \bibfield  {author} {\bibinfo {author} {\bibfnamefont {T.}~\bibnamefont {Chatbipho}}, \bibinfo {author} {\bibfnamefont {R.}~\bibnamefont {Yang}}, \bibinfo {author} {\bibfnamefont {R.~B.}\ \bibnamefont {Wexler}}, \ and\ \bibinfo {author} {\bibfnamefont {L.~B.}\ \bibnamefont {Partay}},\ }\href {\doibase https://doi.org/[doi]} {\enquote {\bibinfo {title} {Adsorbate phase transitions on nanoclusters from nested sampling - raw calculation data},}\ } (\bibinfo {year} {2025})\BibitemShut {NoStop}%
\end{thebibliography}%

\end{document}